\documentclass[sigconf]{acmart}

\usepackage[english]{babel}
\usepackage{blindtext}
\usepackage{amsmath}
\usepackage{enumitem}
\usepackage{booktabs}
\usepackage{multirow}
\usepackage{multicol}
\usepackage{float}
\restylefloat{table}

\copyrightyear{2020}
\acmYear{2020}
\setcopyright{rightsretained}
\acmConference[CoNEXT '20]{The 16th International Conference on
emerging Networking EXperiments and Technologies}{December 1--4,
2020}{Barcelona, Spain}
\acmBooktitle{The 16th International Conference on emerging Networking
EXperiments and Technologies (CoNEXT '20), December 1--4, 2020,
Barcelona, Spain}\acmDOI{10.1145/3386367.3431311}
\acmISBN{978-1-4503-7948-9/20/12}

\newcommand{\point}[1]{\vspace{.05in} \par\noindent\textbf{#1}. }
\newcommand{\subpoint}[1]{\vspace{.05in} \par\noindent\textit{#1}. }
\newcommand{\tool}{\textsc{$\text{CLAP}$}\xspace}

\newcommand{\avgauc}{\underline{\textbf{0.963}}\xspace}

\newcommand{\avgeer}{\underline{\textbf{0.061}}\xspace}
\newcommand{\toponeacc}{\underline{\textbf{76.8\%}}\xspace}
\newcommand{\topthreeacc}{\underline{\textbf{91.0\%}}\xspace}
\newcommand{\topfiveacc}{\underline{\textbf{94.6\%}}\xspace}

\newcommand{\attknum}{73\xspace}

\newcommand{\squishlist}{
        \begin{list}{$\bullet$}
                { \setlength{\itemsep}{0pt}      \setlength{\parsep}{3pt}
                        \setlength{\topsep}{3pt}       \setlength{\partopsep}{0pt}
                        \setlength{\leftmargin}{1.0em} \setlength{\labelwidth}{1em}
                        \setlength{\labelsep}{0.5em} } }
                        
\newcommand{\squishend}{
        \end{list}  }

\begin{document}
\title{You Do (Not) Belong Here: Detecting DPI Evasion Attacks with Context Learning}

\title[You Do (Not) Belong Here: Detecting DPI Evasion Attacks with Context Learning]{You Do (Not) Belong Here: Detecting DPI Evasion Attacks with Context Learning}

\author{Shitong Zhu}
\affiliation{%
  \institution{University of California, Riverside}
}
\email{shitong.zhu@email.ucr.edu}

\author{Shasha Li}
\affiliation{%
  \institution{University of California, Riverside}
}
\email{sli057@ucr.edu}

\author{Zhongjie Wang}
\affiliation{%
  \institution{University of California, Riverside}
}
\email{zwang048@ucr.edu}

\author{Xun Chen}
\affiliation{%
  \institution{Samsung Research America}
}
\email{xun.chen@samsung.com}

\author{Zhiyun Qian}
\affiliation{%
  \institution{University of California, Riverside}
}
\email{zhiyunq@cs.ucr.edu}

\author{Srikanth V. Krishnamurthy}
\affiliation{%
  \institution{University of California, Riverside}
}
\email{krish@cs.ucr.edu}

\author{Kevin S. Chan}
\affiliation{%
  \institution{US Army Research Laboratory}
}
\email{kevin.s.chan.civ@mail.mil}

\author{Ananthram Swami}
\affiliation{%
  \institution{US Army Research Laboratory}
}
\email{ananthram.swami.civ@mail.mil}

\renewcommand{\shortauthors}{Zhu et al.}

\begin{abstract}
As Deep Packet Inspection (DPI) middleboxes become increasingly popular, a spectrum of adversarial attacks have emerged with the goal of evading such middleboxes. Many of these attacks exploit discrepancies between the middlebox network protocol implementations, and the more rigorous/complete versions implemented at end hosts. These evasion attacks largely involve subtle manipulations of packets to cause different behaviours at DPI and end hosts, to cloak malicious network traffic that is otherwise detectable. With recent automated discovery, it has become prohibitively challenging to manually curate rules for detecting these manipulations. In this work, we propose \tool, the first fully-automated, unsupervised ML solution to accurately detect and localize DPI evasion attacks. By learning what we call the \textit{packet context}, which essentially captures inter-relationships  across both (1) different packets in a connection; and (2) different header fields within each packet, from benign traffic traces only, \tool can detect and pinpoint packets that violate the benign packet contexts (which are the ones that are specially crafted for evasion purposes). Our evaluations with 73 state-of-the-art DPI evasion attacks show that \tool achieves an Area Under the Receiver Operating Characteristic Curve (AUC-ROC) of \avgauc, an Equal Error Rate (EER) of only \avgeer in detection, and an accuracy of \topfiveacc in localization. These results suggest that \tool can be a promising tool for thwarting DPI evasion attacks.
\end{abstract}


\maketitle

\section{Introduction}
Deep Packet Inspection (DPI) middleboxes are widely deployed as part of modern network security infrastructures \cite{xu2016survey}. They are stateful , i.e., they not only inspect individual packets, but also reassemble them to form stateful connections defined by a network protocol (e.g., TCP), based on a predefined state machine. To do so, they (for ease of exposition we refer to these as DPIs) need to include a custom network protocol implementation that is often simplified compared to its counterpart on end point platforms (e.g. the OS kernel) due to scarce computation capability, prohibitive overhead and sometimes the need to provide generality in the presence of ambiguous network protocol specifications. 

\point{Adversarial Packets}
Since the advent of DPI, there have been multiple attacks to circumvent them. These attacks  exploit the discrepancies between the DPI's and the OS-level network protocol implementations, to craft subtle yet powerful packets which trigger completely different behaviours on the DPI and at the endpoint (e.g., server). For example, such a packet may be crafted so as to cause the DPI to ignore it, while it still reaches and is accepted by the server. Such packets can potentially contain malicious payloads, and yet successfully bypass the detection of the DPI (because their contents would not be inspected at all). Even worse, prior work shows that in some cases, using only one such packet could cause the DPI to disengage from monitoring of the associated connection, thereby allowing follow-up packets in the connection to pass through without triggering alarms. 

\point{Automated Discovery of Evasion Packets}
In recent years, as network protocol stacks have become increasingly complex with newly added features \cite{ramaiah2010improving}, a growing number of vendor-specific implementations, and continually evolving specifications, there is a surge in research \cite{li2017lib, bock2019geneva, wang2020symtcp} on automating the discovery of \textit{adversarial packets} as described above, to evade DPIs. By applying principled search \cite{li2017lib}, genetic mutation \cite{bock2019geneva}, or symbolic execution \cite{wang2020symtcp}, a vast assortment of adversarial packets can be found with little to no manual intervention. While this works towards understanding attackers, it poses a hard challenge from the defense perspective. As protocol stacks evolve and become more complex, the potential discrepancies that may be discovered with automation can grow in theory, and it becomes prohibitive to manually analyze these subtle implementation issues and patch all of them in the code base; in addition it is hard to generate new hardcoded DPI policies to keep up with new discrepancies that may arise given the pace of rapidly evolving implementations and protocol standards.

\point{Existing Defenses}
Given its difficulty,  only a few limited countermeasures have been proposed against adversarial packets. Notably,  \cite{kreibich2001network}, Kreibich et al.,  apply \textit{traffic normalization} to mitigate the threats of adversarial packets. Specifically, a so-called traffic normalizer is proposed, which acts as a network forwarding element preceding DPI middleboxes, and alters/drops the packets going through the latter as per a predefined set of rules. These rules are  manually curated to describe the "normal traffic" (e.g., the IP header length field must never be smaller or greater than the actual header length). This countermeasure unfortunately, cannot scale in presence of automation -- the achievable search space of all possible adversarial packets can become large enough to make timely manual curation prohibitive. Moreover, the ever-evolving protocol standards introduce false alarms -- a previously correct rule can later cause incorrect decisions (e.g. normal packets being dropped) because of updated implementations. It is also worth noting that as a normalizer, or more generally a \textit{traffic shaper}, \cite{kreibich2001network} provides no detection ability; rather, it blindly alters the traffic stream.

\point{Our Approach}
 Our goal in this paper is to design a practical defense  to effectively detect evasion attempts 
 on DPI middleboxes. 
 Instead of relying on significant manual curation/analysis, we propose a novel, fully-automated (with minimum feature engineering) Machine Learning (ML)-based approach called \tool (\textbf{C}ontext \textbf{L}earning based \textbf{A}dversarial \textbf{P}rotection).
 \tool learns the benign (what we call) \textit{context} from only normal traffic traces (i.e., unsupervised), and uses this learning to detect adversarial packets. In other words, \tool asserts whether the context of the unseen packets ``fit'' in the associated connection or not. 
Specifically, the context of a packet is composed of two types of sub-contexts to describe the aforementioned ``fitness'' of a given packet:
\squishlist
    \item \textbf{Inter-packet context}, which captures the inter-relationships among different packets in the connection in terms of how their header fields change/evolve over the trace; these changes generally relate to the transitioning of states for a stateful network protocol (e.g., TCP);
    \item \textbf{Intra-packet context}, which captures the inter-relationships among different header fields in a given packet (in terms of the combinations of their values).
\squishend
By learning the joint distribution (i.e., the \textit{packet context}) of the two (sub-)contexts from benign traffic, \tool automatically finds violations thereof, caused by adversarial packets.

\point{Motivating Example}
To showcase the intuition behind how \tool works, we present a concrete attack and its detection with \tool. Bad-Checksum-RST is an attack that has been reported in \cite{wang2020symtcp, bock2019geneva} and shown to be effective against Great Fire Wall (GFW), a state-of-the-art DPI-based censorship system. It injects an ill-formed RST packet with a garbled TCP checksum value after the three-way handshake. Since common endhost TCP implementations perform a rigorous checksum verification but the GFW does not, this injected RST packet is dropped by the endhost but not GFW. As a result, upon seeing a RST packet, GFW would disengage its monitoring of the connection, while the communications between two endhosts would be allowed to continue. 


By learning the benign context from a large set of clean network traces, \tool knows
what requirements (inter- and intra-packet contexts) must be met by a packet at its position in the connection (e.g., can it be a RST and if so, should its checksum be correct?); then, \tool checks whether the  packet conforms (fits in) to a benign packet context. In this case, the RST packet with a bad checksum value that appears after three-way handshake is  asserted as violating both the inter- (RST should not take place at this point) and intra-packet (checksum of RST packet should be correct) contexts (based on training) and thus, the evasion attempt is detected. 


\point{Contributions}
In brief, our contributions in this paper are:
\begin{enumerate}
    \item We are the first to propose a fully-automated unsupervised learning approach to detect adversarial packets that are crafted to elude DPI middleboxes.
    \item Our evaluations on 73 state-of-the-art DPI evasion attacks over realistic backbone traffic captures, show that by only learning from benign traffic, \tool achieves an overall detection AUC-ROC of over \avgauc, with an average EER of only \avgeer (the two most commonly used evaluation metrics for ML-based IDSs \cite{mirsky2018kitsune,aqil2017jaal,marin2018rawpower,marin2019deepsec}).
    \item Our evaluations show that beyond detection, \tool achieves an average Top-5 localization accuracy (identifying a train of five packets most likely to contain the attack vector) of \topfiveacc (Top-3 of \topthreeacc).
    \item Our performance analysis shows that our pipeline can process over 2,100 packets per second on a single CPU core, with linear scalability. 
    \item We will open source both our implementation of \tool, and the used datasets, at https://github.com/seclab-ucr/CLAP to allow reproducibility and future extensions.
\end{enumerate}
\section{Related Work}
\label{sec:related_work}
\point{ML-based Intrusion Detection System (IDS)}
As a critical and commonly deployed network infrastructure, intrusion detection systems (IDSs) are designed to detect suspicious traffic and flag them accordingly. Traditionally, IDSs are signature-based and catch malicious traffic that violates a predefined set of rules. This type of IDSs face challenges because their manually curated signatures cannot keep up with emerging threats. In contrast, anomaly-based IDSs do not rely on priori-created signatures; an anomaly-based IDS inspects and classifies traffic by asserting whether it aligns with  patterns observed in normal traffic. More recently, ML techniques have been used for anomaly detection, as they are efficient in learning patterns from benign traffic and then distinguishing between normal and suspicious instances on unseen traffic \cite{mirsky2018kitsune,marin2018rawpower,marin2019deepsec,aqil2017jaal}. 
While with a similar general goal (detecting suspicious traffic), \tool is fundamentally different because it (1) spots attacks that specifically seek to evade DPI detection, not for general malicious purposes (e.g. DDoS); and (2) uniquely considers packet context that is critical for detecting DPI evasion attacks, while context-agnostic ML-based IDS is unable to do so as will be shown in our evaluations.


\cite{jingping2019detection} is the only attempt towards using ML to discover DPI evasion attacks to our best knowledge. However, the approach proposed in \cite{jingping2019detection} relies on training its model on both benign and malicious traffic traces, which renders a different threat model as compared to what we assume in this paper, and is therefore not comparable directly. In fact, one of the key attributes of \tool is its ability to detect subtle evasion attacks without a priori knowing about them.
For an apples-to-apples comparison, we use a state-of-the-art unsupervised IDS \cite{mirsky2018kitsune} as one of the baselines in Section \ref{sec:eval}, and show that the state of the art general-purpose IDS is ineffective/inaccurate in detecting DPI evasion attacks because it does not consider context.


\point{Context Learning}
The general notion of using context has been explored in computer vision for improving classifier performance \cite{tang2019learning,liu2018structure}. Context refers to co-occurrences of objects that commonly appear together in the same scene, and aids detection/segmentation tasks where certain objects lack inherent patterns to be recognized, but can be inferred by occurrences of other easily-recognizable objects. Note that while these approaches, by mainly characterizing spacial inter-relationships among different objects as context, are generally appropriate for vision applications (i.e., objects in same scene), it cannot be applied directly to network domain applications. Context inconsistency has also been very recently considered for thwarting adversarial examples in computer vision \cite{li2020connecting} but has never before considered in the network intrusion detection context. Remotely inspired by these efforts, for the first time, we define {\it packet context} for network traffic data, and propose \tool to automatically learn and apply this to detect DPI evasion attacks. We consider the unique characteristics of network traffic, and design our system accordingly. 


\section{System Design}
\subsection{Overview}

\begin{figure}
    \centering
    \includegraphics[width=\columnwidth]{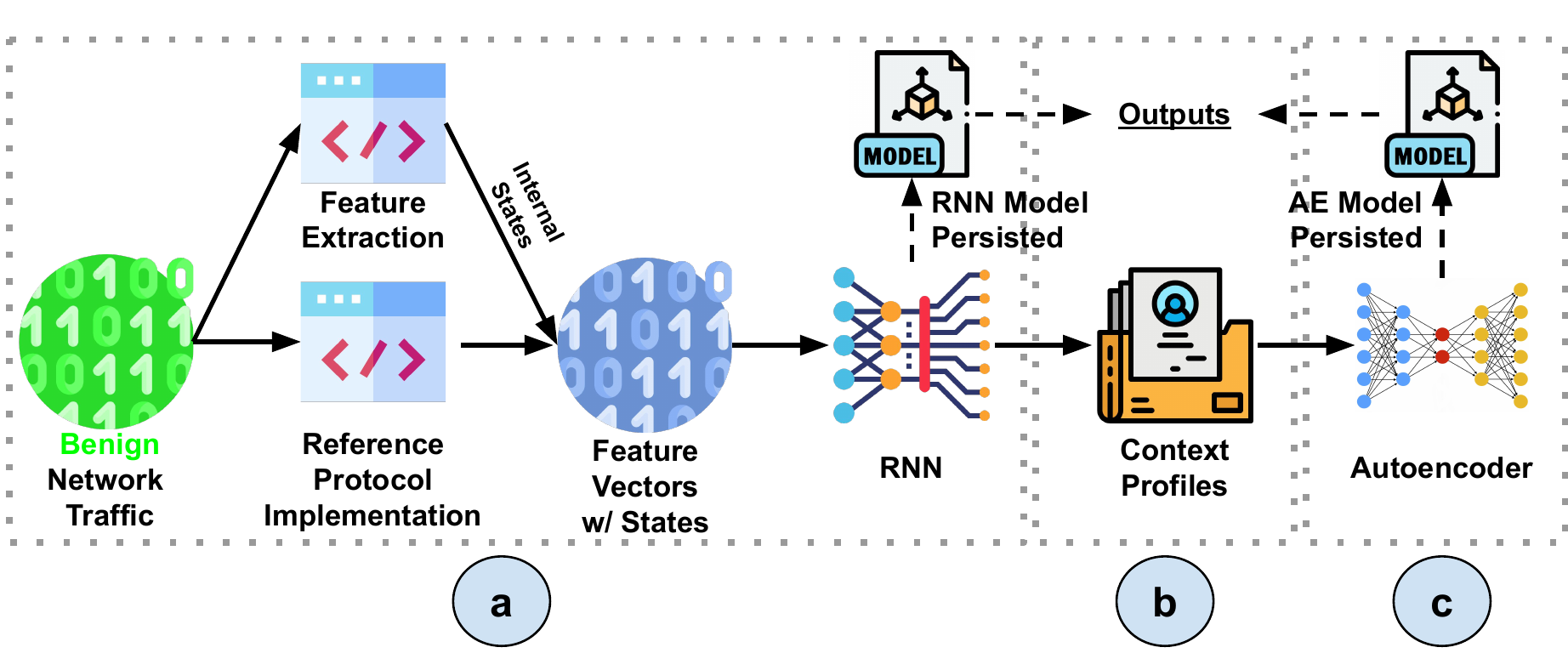}
    \caption{Training phase of \tool}
    \label{fig:training_phase_achitecture}
\end{figure}

\begin{figure}
    \centering
    \includegraphics[width=\columnwidth]{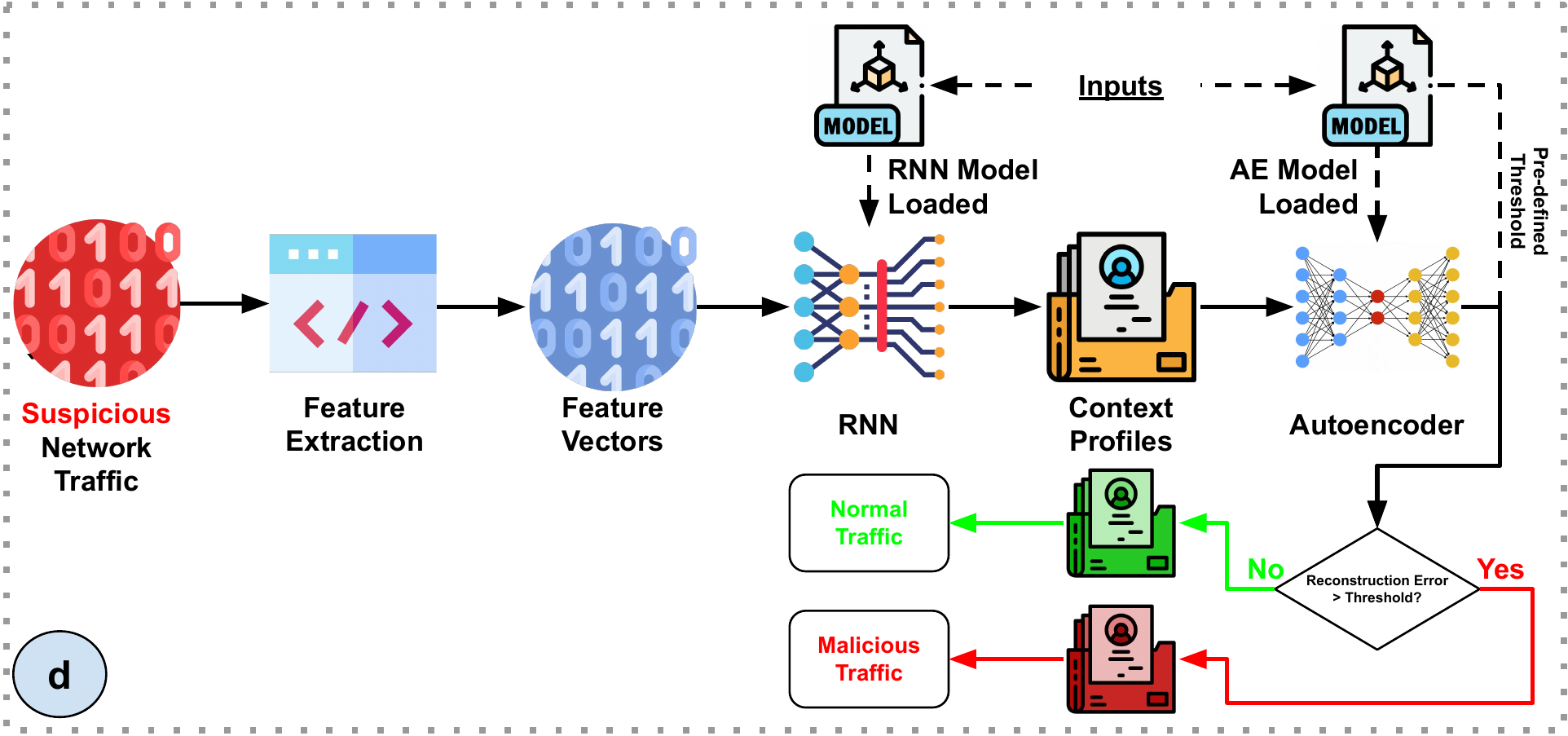}
    \caption{Testing phase of \tool}
    \label{fig:testing_phase_achitecture}
\end{figure}

As discussed earlier, \tool draws on the fact that there are co-occurence relationships
between packet fields (intra-packet), and across packets (inter-packet) in a single TCP flow. To re-iterate, these relationships form the packet context. An evasion attack is likely to tamper with these
relationships to confuse the DPI middlebox  
 and \tool seeks to 
 most efficiently learn and use the packet context to detect such tampering. Our design of \tool consists of 4 stages: 
\squishlist 
    \item (a) Learning benign inter-packet context by training a RNN model whose task is to predict the transitions across a set of connection states (not only high-level TCP states but also subtle verdicts/states, as described later), on benign traffic traces; 
    \item (b) Fusing/concatenating the benign inter-packet context (i.e., as discussed later gate weights from (a)'s RNN model represent inter-relationships among different packets) and intra-packet context (i.e., combinations of packet header fields) to generate benign \textit{context profiles}; 
    \item (c) Learning benign holistic packet context by characterizing the distribution of context profiles generated in (b); 
    \item (d) Detecting DPI evasion attacks by verifying whether the context profile of unseen packet trains violates the distribution of benign context profiles as learned in (c).
\squishend
We show diagrams that depict Stage (a)/(b)/(c) of \tool in Figure \ref{fig:training_phase_achitecture}, and (d) in Figure \ref{fig:testing_phase_achitecture}.

\begin{figure}
    \centering
    \includegraphics[width=0.8\columnwidth]{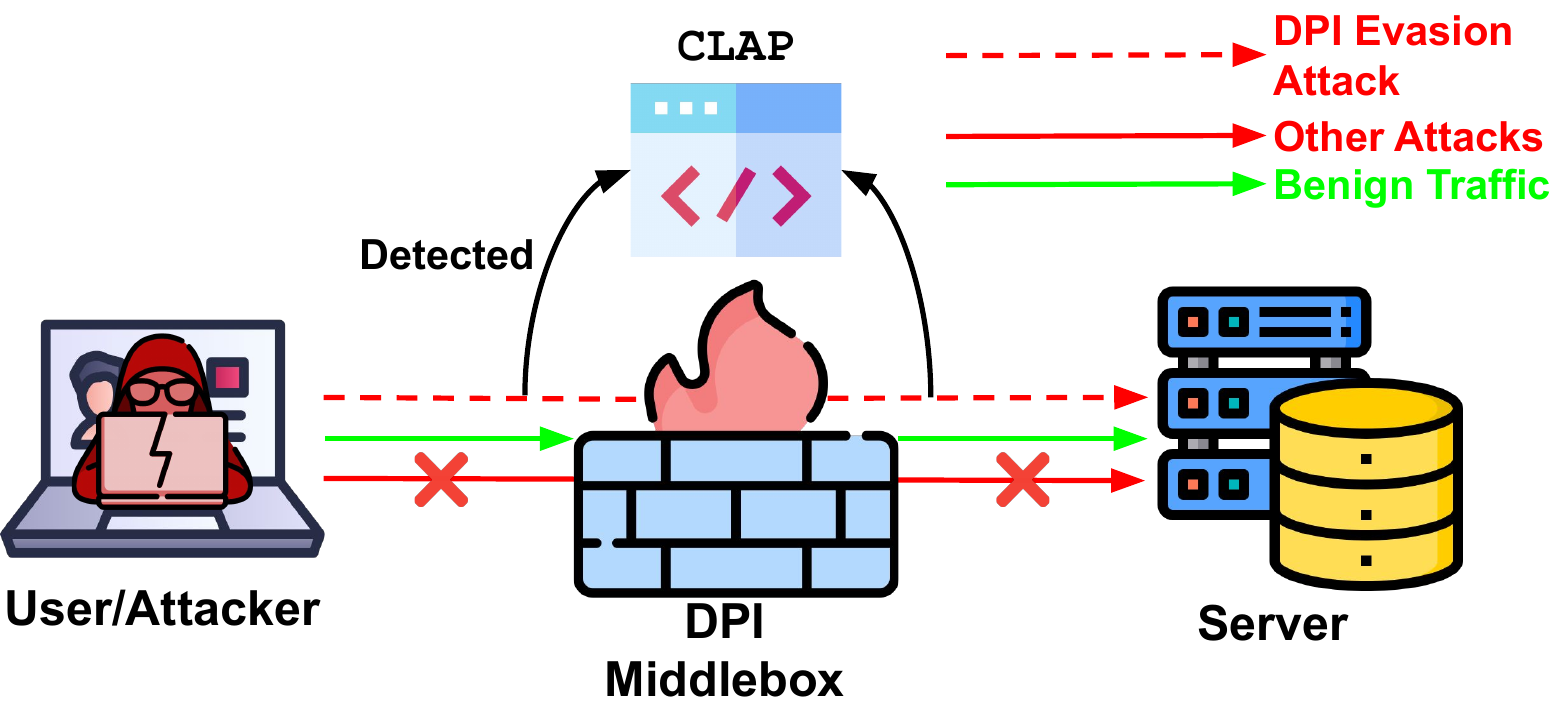}
    \caption{Threat model of DPI with \tool}
    \label{fig:threat_model}
\end{figure}

\subsection{Threat Model}
Before diving into the technical details of the design of \tool, we establish the threat model we consider. As previously mentioned, \tool is designed to protect stateful DPI middleboxes against evasion attacks. We capture the threat model associated with common DPI middleboxes relating to our work, and the role of \tool in Figure \ref{fig:threat_model}. We assume both the DPI middlebox and \tool are located in between the client and server, and capable of reading all packets going through. Commonly, DPI inspects payloads of these packets and detects malicious contents in those payloads (i.e., "Other Attacks" in Figure \ref{fig:threat_model}). When the attacker launches the DPI evasion attack from client side, he/she injects specially crafted packets (i.e., adversarial packets) to cause discrepancies in behaviours between the DPI and the server. The discrepancies cause the follow-up malicious traffic to not be inspected by the DPI (but can still reach the server). \tool is designed to specifically detect such adversarial packets. Note that we limit the scope of our target evasion attacks to those that only involve header fields manipulations, due to (1) scarcity of payload-related evasion strategies; and (2) prohibitive overheads in training for benign payloads. We only focus on the TCP protocol in this work since it is arguably the most popular transport layer protocol and is most commonly targeted in DPI evasion attacks. Since \tool does not depend on the DPI itself, it can function interdependently (for forensic analysis)
and does not rely on anything besides raw traffic that traverses the DPI middlebox (i.e., PCAP captures). Hence, \tool can not only be deployed as an online detector complementing existing DPIs to detect evasion attacks (with affordable overhead as shown in Section \ref{sec:eval_perf}), but also be used as a forensic tool to analyze the traffic captures offline.


\subsection{\tool Design}
Next, we describe the different components in \tool that help achieve its overarching
goal.

\point{(a) Learning Inter-packet Context}
\label{subsec:train}

\subpoint{Goal}
The first stage of \tool involves training a RNN model to drive it towards learning the inter-packet context as defined previously. As established earlier, the inter-packet context  essentially captures the relationships among the header fields (co-occurrence) in a train of packets. We reiterate that this context strictly describes the relations across \textit{different} packets, and therefore does not concern itself with any relations/combinations of header fields within the same packet. 

In order to capture these inter-packet relationships with a ML model, we need to design a corresponding learning task. 
Since the transitions of a TCP state machine depend on such relationships across a sequence of given packets, 
we use a ML classifier that predicts these transitions; such a classifier will need to
learn the temporal structure across packets. 
The best option among various neural network choices for doing so is a GRU-based RNN model. Based on this general architecture, we design a customized model that takes the header fields of each packet as inputs, and predicts the corresponding states to which the reference TCP state machine will transition as a result of this packet, as model output. Note that we are not interested in the classification result of the RNN, but rather require the distribution of the benign inter-packet contexts which we draw from the weights of the gates in the neural network architecture, as discussed in the next subsection. This requires  exceptionally high performance for the task we design for RNN (connection state prediction), which from our evaluations (detailed in Table \ref{tab:rnn_accuracy}), is achieved with an overall prediction accuracy over 0.99).

\subpoint{RNN/GRU}
RNNs are a class of neural networks that are widely used to model temporal data. In brief, they contain a recurrent cell that processes one element in the input sequence a time, considers both the current input and a memory state from the previous unit to output a new memory state. Beyond accomplishing classification tasks, the chaining of these repeated units in RNN models have also been highly successful \cite{tang2019learning,liu2018structure} in generally modeling and encoding the interrelationships across the input sequence. Among RNN cells, the Long Short-Term Memory Unit (LSTM) and the Gated Recurrent Unit (GRU) are considered the state-of-the-art architectures. Both of them consist of a "gating" mechanism, wherein gates that are in the cells update weights that represent the relationships across input sequence. We pick GRU in this work as it provides performance that is comparable with LSTM but incurs considerably lower overhead, and note that LSTM is also a viable option.

\subpoint{Input Feature Set}
In order to cover (1) all header fields that influence the transitions of the TCP state machine
and (2) header fields that can possibly be manipulated by DPI evasion strategies (i.e., all non-tuple-related, non-optional IP/TCP headers, and a few common TCP option headers, as specified in \cite{postel1981rfc0791,rfc7931981transmission}), 
we include 37 header field values as features (7 for IP and 25 for TCP, full list in Table \ref{tab:feature_set}) from the IP and TCP headers. To reiterate, we do not consider payload-related features here because (1) they are irrelevant to TCP state transitions; (2) there are only a very small fraction (7 out of 80 from \cite{wang2020symtcp,bock2019geneva,li2017lib}) of DPI evasion attacks that involve manipulating packet payloads; (3) it is prohibitively challenging to learn distributions of unstructured data such as benign payloads; we leave this as an open question to examine in the future; and (4) most public traffic archives (including what we use in the evaluations) strip payloads in the traces for privacy concerns.
We alto follow the general principle of using these fields in the raw form to the extent possible to avoid heavy feature engineering (i.e., only need minimum pre-processing such as validating checksum, as detailed in Table \ref{tab:feature_set}), and show that the RNN is capable of inferring the connection states without requiring extensive domain knowledge. 

\subpoint{Labeling} According to the IETF standard \cite{rfc7931981transmission}, TCP defines 11 different states (e.g. \texttt{SYN\_SENT/SYN\_RECV/ESTABLISHED}) that we refer to as \textit{master TCP states}. In addition to these states, in order to enrich the learned packet context, especially in the rather coarse grained ESTABLISHED state, we also include the more subtle, in-/out-of-window states (i.e., whether an incoming packet is within the recipient's receive window) collected from the reference TCP implementation to be part of the label. Therefore, the label we use to train RNN is the concatenation of the master TCP state and the in-/out-of-window subtle \textit{state} \footnote{Although, we are slightly misusing the term, when we refer to state from hereon, we not only include the traditionally defined TCP states, but also a packet classification/verdict (in-window or out-of-window) that strongly relates to inter-packet relationships, to better drive the RNN model to learn bengin inter-packet contexts.}, resulting in a total of 11 * 2 = 22 potential classes (an in-/out-of-window possibility is included for each of the 11 master states); these are listed in Table \ref{tab:feature_set}. In order to collect reliable and accurate states for labeling the training data of our RNN, we resort to OS-level (e.g. Linux) TCP stack implementations and instrument their relevant modules to expose our desirable states (i.e., master TCP state and in/out-of-window subtle state). We replay the benign training traffic captured on the instrumented platform to harvest their corresponding states as labels (the details of the setup are in Section \ref{sec:eval_setup}).

\subpoint{Training}
To put things together, we now formally describe how we train the RNN model. Consider a benign network connection consisting of $n$ packets $P_{1...n}$, where  $P_i=[F_{IP}^{1},...,F_{IP}^{8},F_{TCP}^{1},...,F_{TCP}^{29}]$ (i.e., $F_{IP}^{i}$ represent the features from the corresponding packet's IP header fields, and $F_{TCP}^{i}$, those from the TCP header fields). We train a RNN model $M_{GRU}$, with GRU as its cell architecture, by feeding $P_{1...n}$ and their corresponding ground-truth labels (i.e., states from TCP state machine) $L_{1...n}$, and executing standard error back-propagation \cite{cho2014learning}.
The standard multi-class cross entropy loss function in Equation \ref{eq:loss},
where $L_i$ and $\hat{L_i}$ are the probabilities relating to class $i$ of the ground-truth label vector and the predicted label vector, respectively, is used as the loss function for training the RNN. It measures the error between the current RNN model output (i.e., a vector of probabilities) and the ground-truth label (i.e., a vector of values of 0 except that the index of expected/ground-truth state is 1). The error is propagated back to the prior layers of the RNN to update its parameters until the model produces the correct output.
In the next subsection, we describe how the intermediate/latent gate states of the training GRU are used to build context profiles. 
\begin{equation}
\begin{split}
    Loss_{CrossEntropy}(L, \hat{L}) &= -\sum_i {L_i log(\hat{L}_i)}, \\
            \textrm{where } \hat{L} &= M_{GRU}(P)
\end{split}
\label{eq:loss}
\end{equation}


\point{(b) Fusing Inter- and Intra-packet Contexts}

\subpoint{Goal}
Once the RNN model is trained, its gates contain the inter-packet context learned from benign traffic traces (described in more detail in the next paragraph). Next, we need to extract the intra-packet context and fuse it with the inter-packet context to form the \textit{context profiles} of a packet. Recall that the intra-packet context is defined by the relationship/combinations across header field features within the packet. We fuse the two contexts by simply concatenating them (i.e., gate weights and header field values) into a unified feature vector which is referred to as the context profile for that packet. This fusion strategy enables the autoencoder in Stage (c) (to be described) to learn the joint distribution of both contexts. By examining this joint distribution, \tool is capable of exposing attacks that violate (1) only the inter-packet context; (2) only the intra-packet context; and (3) both sub-contexts simultaneously.

\begin{figure}
    \centering
    \includegraphics[width=0.85\columnwidth]{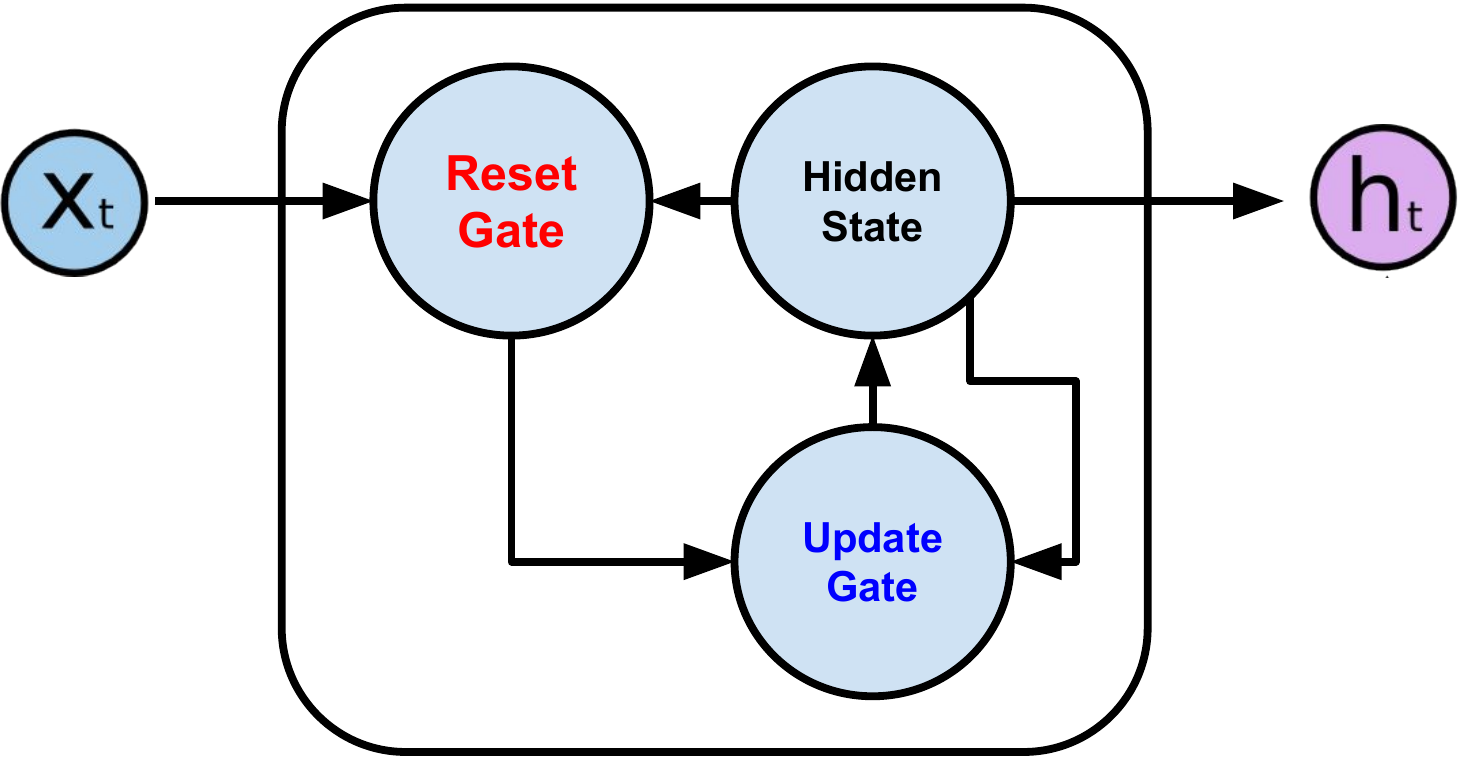}
    \caption{Internals of GRU cell}
    \label{fig:GRU_internals}
\end{figure}

\subpoint{Gate Weights}
Figure \ref{fig:GRU_internals} provides a closer look at the internals of GRU cells. For one such cell, besides the input (i.e., $x_t$ in Figure \ref{fig:GRU_internals}) and the output state (i.e., $h_t$ in Figure \ref{fig:GRU_internals}), there are also "gates" for optionally letting information through \cite{cho2014learning} (i.e., reset and update gates as marked in Figure \ref{fig:GRU_internals}). 
In order words, these gates explicitly control whether the output classification (i.e., the TCP state described earlier) of a given packet strongly relates to its previous packets, or not. Specifically, if at the current time step $t1$ the gate weights with respect to a previous packet $P_{t2}$ at time step $t2$ are large, it means the features of $P_{t2}$ contribute greatly to the classification of the next output at $t1$ (and vice versa). As one can see, the gate weights are ideal means to characterize the dependencies or inter-relationships across the different packets in a connection, or in other words, capture the inter-packet context.

\begin{figure}
    \centering
    \includegraphics[width=0.85\columnwidth]{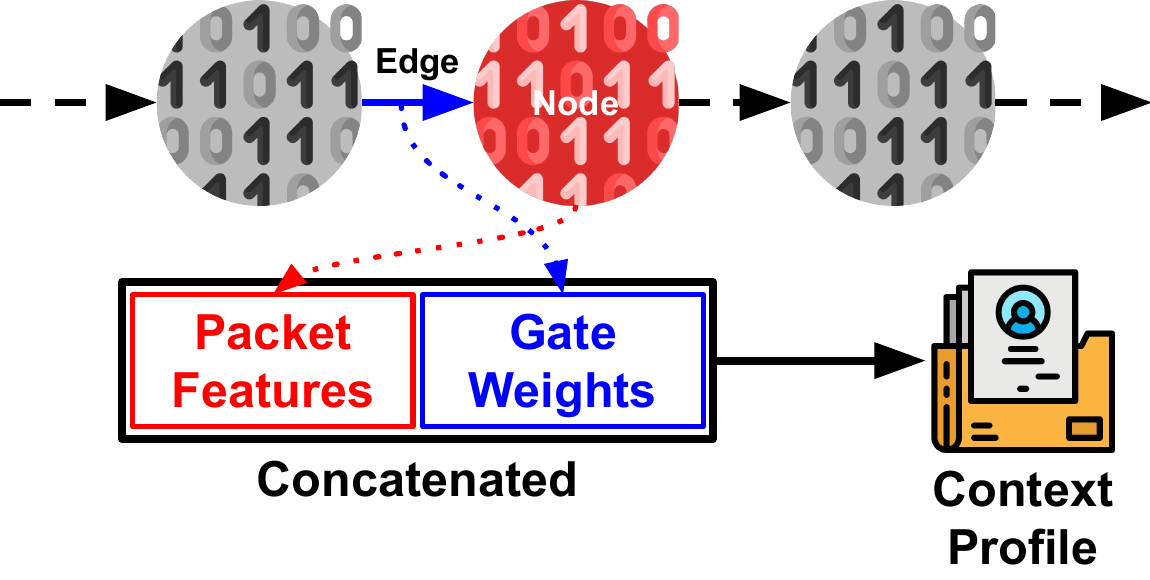}
    \caption{Representation of context profile as chain graph}
    \label{fig:chain_graph}
\end{figure}

\subpoint{Chain Graph}
To visualize the context profile of a packet, we can represent the same as a graph-based model wherein the packet header features are the graph nodes, and gate weights are the edges connecting packets (nodes). Thus, the train of packets can be represented as a chain-shaped graph. The packets header field features are the node features, and the edges between two adjacent nodes (consecutive packets) are the GRU gates that describe the inter-relationship between the two packets. The resulting connected chain graph, models the network trace. 
The gate weights (edges) in the graph, (which are learnt during training) control the information propagation between the adjacent packets (nodes). We depict this chain like graph model in Figure \ref{fig:chain_graph}.

\subpoint{Amplification Features}
While we have included all necessary packet header fields as the feature set for learning the intra-packet context, we discover that some subtle context violations are challenging to capture in practice with just these features. These relate to extremely small perturbations in terms of the associated changes of certain features; these help in successfully evading the DPI, but without amplification seem too insignificant for the ML classifier (specifically the autoencoder) to  recognize. To address this, we augment our feature set with two types of \textit{amplification features}, that are incorporated into the context profile; these features, crafted based on domain knowledge, amplify the aforementioned subtle context violations such that the distribution discrepancy they cause are easily captured by the autoencoder that is described later when we discuss  Stage (c). In particular, the amplification features that we design are: (1) \textit{out-of-range features}, which indicate whether the packet's associated numerical header field value is out of the range of what has been observed in the benign training traces; and the (2) \textit{equivalence relation feature}, which indicates whether an expected equivalence relationship with respect to certain header field values is maintained (e.g., TCP payload length = IP total length - IP header length - TCP data offset) 
or not. We provide the full list of the 15 amplification features (all belonging to the two types) with their semantics in Table \ref{tab:feature_set}. We refer to the concatenation of the raw header field value features and the amplification features as \textit{packet features}.

\subpoint{Concatenation}
We next provide details on ``how exactly'' we generate the context profile. Consider a GRU taking $n$ input features with its hidden state size also being  $n$-dimensional; in such a case, the size of its gates should be equal to the hidden state (i.e., $n$) \cite{cho2014learning}. 
As previously discussed, the context profile is the concatenation of the packet features and the gate weights, as defined in Equation \ref{eq:context_profile_def}. We provide a full list of all packet features in Table \ref{tab:feature_set}. 
Upon generating the context profiles for all the packets in a train, we can concatenate consecutive packet profiles to form a ``stacked profile.'' This design aggregates multiple context profiles from consecutive packets, and therefore explicitly embeds inter-packet relationships into the stacked profile (in addition to existing gate weights that are also designed to capture inter-packet context); this in turn provides profiles with richer inter-packet context encoded. As one might expect, we find that this helps improve performance and eventually use it in our evaluations.
\begin{equation}
\begin{split}
           CxtProf &= [P_{IP}, P_{TCP}, P_{amp}, G_{forget}, G_{update}] \\
    StackedCxtProf &= [...,CxtProf_{t-1},CxtProf_t,CxtProf_{t+1},...]
\end{split}
\label{eq:context_profile_def}
\end{equation}
\point{(c) Learning the Joint (Context) Distribution}

\subpoint{Goal}
Once the benign context profiles that contain both inter- and intra-packet contexts are generated as described in (b), \tool needs to learn their joint distribution so that it can later detect suspicious packets that violate the benign context (i.e., joint distribution). 


\subpoint{Autoencoder}
Antoencoders are a class of neural networks that 
characterize the distribution of given training samples by forcing the networks to reproduce the inputs themselves as model outputs. 
In other words, they are tasked to encode a given input to a compressed latent space (bottleneck layer), and decode it to recover the input as much as possible.
During this process, the autoencoder learns the compressed representation from the training data (i.e., benign context profiles in our case) distribution as the features of its bottleneck layer, and the reconstruction error between the real input and recovered input (model output) is considered an ideal metric to characterize an input (context profiles) in terms of how close it is to the learned distribution. 
If a context profile 
traversing an autoencoder trained on benign context profiles, exhibits a high reconstruction error, we infer that it deviates from, or violates the benign context. 
Autoencoders are considered as the state-of-the-art means for anomaly-detection tasks \cite{zhou2017anomaly}.
We use an autoencoder here to learn the distribution of benign context profiles, and then determine if the context of unseen packets is consistent with what is observed in benign packet traces in (d).  


\subpoint{Training}
We train the autoencoder with benign context profiles obtained.
The L1 loss function is used to measure the reconstruction error and is the sum of the all the absolute differences between the true value (ground-truth context profile) and the predicted value (reconstructed context profile). During
training, by minimizing the L1 loss, the autoencoder learns to characterize the distribution of benign context profiles. The choice of the L1 loss function also relates to its excellent performance 
in handling dense input data, meaning data wherein there is little to no sparsity (i.e., features with zeroes as values). This suits the properties of the context profiles generated in Stage (b), because both the packet features and the gate weights are dense and rarely have 0 as values.
\begin{equation}
\begin{split}
           Loss_{L1}(X_{input}, X_{output}) &= \frac{1}{n}\sum_{i=1}^n |X_{output}^i - X_{input}^i| \\
           \textrm{where } &\{X_{input}, X_{output}\} \in \mathcal{R}^n
\end{split}
\label{eq:l1_loss}
\end{equation}


\point{(d) Verification}
\subpoint{Goal}
Lastly, \tool with its trained RNN (to generate gate weights for context profiles) and autoencoder (trained using those profiles), is to be deployed online (i.e., it encounters previously unseen packets) to detect possible DPI evasion attacks. Recall that the autoencoder from Stage (c) can only compute the reconstruction error with respect to each context profile. Given that the input to \tool would be connections/sequences of packets and our goal is to provide a connection-level detection conclusion, we need to first determine a strategy for \tool to compute a score (which captures the likelihood of whether there are evasion packets within the connection for a given connection that contains sequence of reconstruction errors produced by the trained autoencoder. Then, \tool must analyze the context profiles for unseen packets (their conformance to benign profiles) and use the chosen score (discussed below) to assert if this connection is adversarial. 

\subpoint{Adversarial Score}
\label{sec:adv_score}
Once the autoencoder is trained, it produces a numeric reconstruction error for a given (stacked or non-stacked) context profile. Consider a connection that consists of $n$ packets in total. Let us assume that  stacked context profiles containing $t$ packets, are generated in a sliding-window fashion (i.e., concatenation of every $t$ consecutive single-packet profiles from the beginning to the end of a unidirectional traffic sequence; these profiles are overlapping) are obtained. Specifically, with this approach, \tool  generates $n-t+1$ such stacked profiles (i.e., leading to $n-t+1$ tests and thereby, reconstruction errors at the autoencoder stage) for the connection. This sliding window enumerates all possibilities of temporal contexts, and can be expected to provide the best coverage. \tool now needs to capture a "summarized" value that characterizes the "overall profile" of the connection, by means of  these enumerated profiles. There is a spectrum of approaches for characterizing a group of observations in general, ranging from basic statistical quantities such as maximum/minimum, variance, mean, median, to more advanced methods such as training a separate autoencoder for the summarization \cite{mendenhall2012introduction}.

\begin{figure}
    \centering
    \includegraphics[width=\columnwidth]{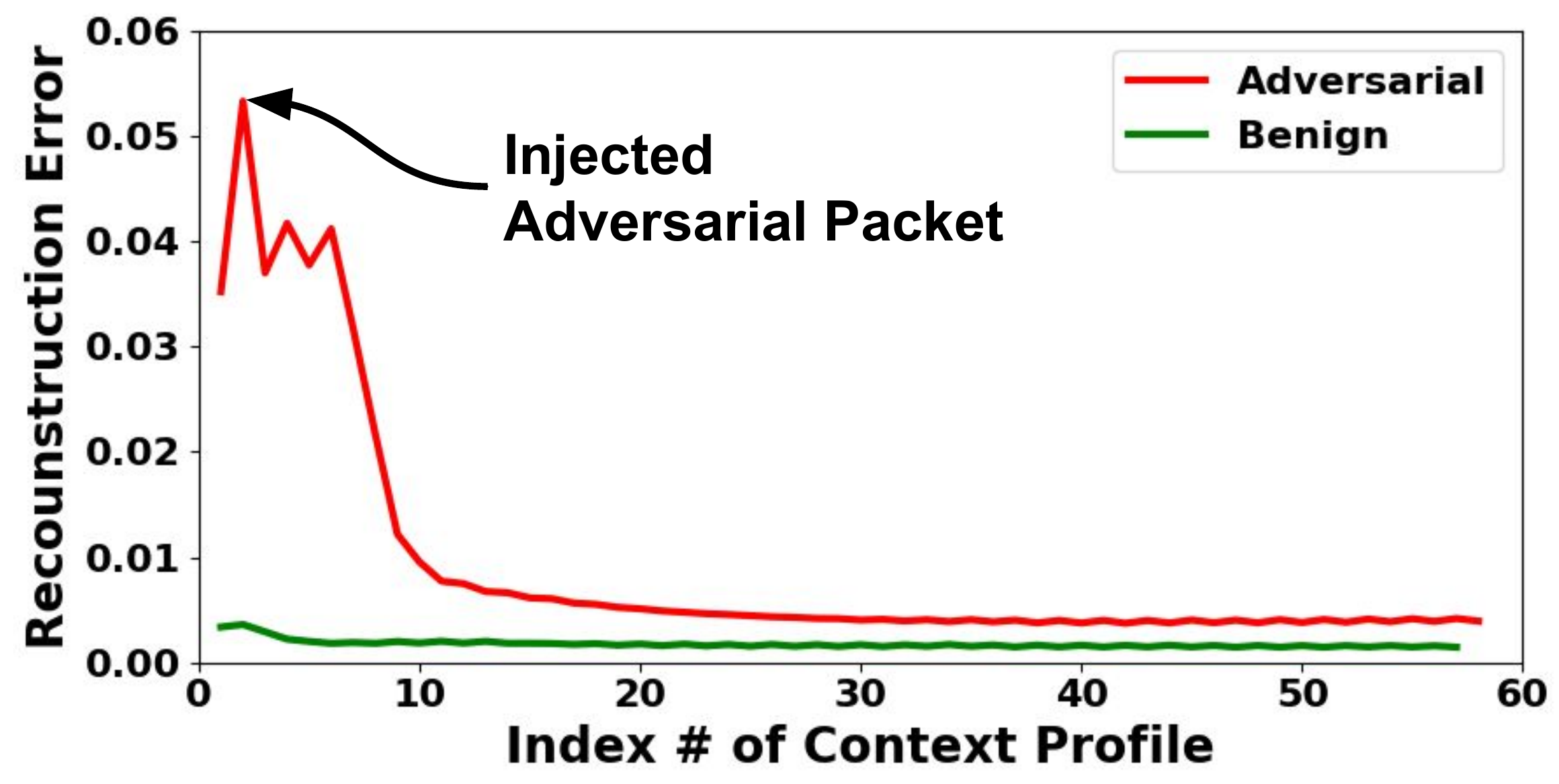}
    \caption{Typical trend of reconstruction errors across a connection}
    \label{fig:recon_err_trend}
\end{figure}

We examine the  reconstruction error trends from adversarial connections, towards empirically choosing a proper metric. An example is shown in Figure \ref{fig:recon_err_trend}, where an adversarial packet introduces a spike in the error value (around the time it is encountered) and then, the error level falls off to get closer to that with benign profiles.
Based on this observation (which is expected since the sliding windows closest to the adversarial packets are likely to show the highest error), we propose a  \textit{localize-and-estimate} approach to choose a metric; this maximizes our odds of distinguishing adversarial from benign connections in the testing phase. Specifically, we first localize (identify the position within the sliding window) the profile with the maximum reconstruction error in the connection; and (2) sample the mean reconstruction error over the window (of 5 profiles in this work) by choosing the profile with the maximum reconstruction error as center. We refer to this mean as the \textit{adversarial score} for the connection. We believe this newly designed metric best captures the most distinguishing (the spike) part of the reconstruction error sequence in a connection, provides the best detection performance. 
In addition to the score, to make a Boolean classification (attack or no attack), we also need a threshold to determine at what level of the adversarial score,  do we consider the connection to be an attack. The choice of this threshold will provide a trade-off between true and false positive rates. We leave the freedom of choosing the threshold to the deployer of \tool towards achieving the appropriate trade-off;
however, our results in Section~\ref{sec:eval_eff} offer possible choices for these thresholds (by means of ROC curves). 


\subpoint{Deployment}
Lastly, with the 3-tuple $\{M_{GRU},M_{AE},TH_{Adv}\}$ (i.e., trained GRU-based RNN model, trained autoencoder model and selected threshold for adversarial score) generated in the previous steps and an incoming (suspicious) connection $C_{Sus}$, the deployer of \tool first executes the pipeline in Figure \ref{fig:testing_phase_achitecture} to compute the adversarial score for $C_{Sus}$, and then compares it with $TH_{Adv}$ to determine if the connection is adversarial or not. 
Note that beyond binary classification (i.e., adversarial or benign), \tool can also pinpoint the most "suspicious" packets in a connection, by localizing the ones that bear the highest reconstruction error (i.e., the first step of "localize-and-estimate" approach). We will evaluate both the detection and localization accuracies in the next section.

\section{Evaluations}
\label{sec:eval}
In this section, we provide comprehensive evaluation results with regards to different aspects of \tool, viz.,: (1) effectiveness analysis in terms of how well it detects and localizes adversarial packets, compared to other baselines; (2) case studies with in-depth analysis on its performance on certain attacks; and (3) performance analysis in terms of how fast it can process network traffic in practice. 

\subsection{Setup}
\label{sec:eval_setup}
\point{Dataset}
We first describe the dataset used in our evaluations. For learning benign packet contexts that are as complete and sound as possible, the dataset of benign traffic traces must guarantee that it (1) only consists of benign (i.e., no DPI evasion attempts) flows; and (2) is adequately representative i.e., non-adversarial packets including those with uncommon but legitimate patterns are well-covered. Given these requirements, we select the MAWI Traffic Archive \cite{MAWIdataset} that provides PCAP captures of a backbone network in Japan. MAWI traffic traces are considered one of the state-of-the-art long-term measurements on wide-area/global Internet and have been used in prior networking/security research \cite{cho2017recursive,fontugne2017scaling,aqil2017jaal}. 
Note that the payloads of MAWI traces (and most other public datasets) are all stripped (payloads are not visible) for privacy concerns, making it impossible to use for detecting payload-related attacks (e.g. segmentation-based attacks) -- we adjust the corpus of evaluated attacks accordingly. Table \ref{tab:dataset} (in Appendix due to space limitation) lists basic statistics of the traffic capture we use.


Recall that Stage 1 of \tool needs a reference TCP implementation to generate the internal states required for training the RNN model. We select the \texttt{conntrack-tools} module \cite{ayuso2008conntrack} in Linux's \texttt{Netfilter} sub-system that provides standard and reliable infrastructures for packet inspection. We instrument relevant code in Linux kernel version 5.6.3 to expose the required states, and use it as the traffic ``replayer'' for collecting labels for RNN training.

\point{Simulated Attacks}
For generating the the traffic that seeks to evade the DPI middlebox, and to include these to form the dataset, we adopt a simulation-based approach. We integrate these traffic flows into the benign MAWI traffic traces. 
Specifically, we thoroughly analyze the evaluated 73 DPI evasion attacks proposed in \cite{bock2019geneva,wang2020symtcp,li2017lib}, and implement a simulator that injects the modifications incorporated in these to evade the DPI middlebox, into the MAWI traffic traces (i.e., PCAP files).
We acknowledge that this PCAP-level simulation might potentially introduce some slight divergence in behaviors, compared to what happens in live DPI evasion attacks (e.g. the timings associated with the injected packets is estimated in our simulation, which could differ from the live case due to unpredictable congestion effects). However, we argue that (1) these differences do not disrupt the underlying mechanisms of DPI evasion attacks and thus, do not affect our evaluation results/conclusions; and (2) the open-source implementations released by \cite{bock2019geneva,wang2020symtcp,li2017lib} do not support a direct replay of the attack traces and, thus we are unable to verify those directly; however, the simulations are a faithful reproduction of those attack behaviors and we assume them to yield similar success against DPI middleboxes.

\point{Baselines}
We compare \tool's performance with that of 2 baselines to showcase its effectiveness. Baseline \#1 reuses the same pipeline as \tool, but (1) removes all gate weight features from its context profiles, (2) limits the length of the profile to single packet. In other words, only intra-packet context features are considered to eliminate inter-packet context information. One can think of this as a ``temporal context'' agnostic version of \tool.
Baseline \#2 is the faithful reproduction of a state-of-the-art anomaly-detection-based IDS \cite{mirsky2018kitsune}. This IDS also leverages autoencoders as its underlying model and the paper claims that it targets a broad spectrum of attacks. We will show the results with these baselines, in comparison with those with \tool's in Figures \ref{fig:res_symtcp}, \ref{fig:res_liberate} and \ref{fig:res_geneva}.

\subsection{Effectiveness Analysis}
\label{sec:eval_eff}
\textit{Takeaway}: \tool achieves an average AUC-ROC score of \avgauc (vs. 0.846 [-12.1\%] for Baseline \#1 and 0.498 [-48.3\%] for \#2), Equal Error Rate (EER) of \avgeer (vs. 0.198 [+224.6\%] for Baseline \#1 and 0.502 [+723.0\%] for \#2) in detecting attacks; it also achieves a \topfiveacc Top-5 (meaning the localized packet is within a window of five packets from the identified point), \topthreeacc Top-3 (the localized packet is within a window of 3 packets), and \toponeacc Top-1 (exactly identify the position of the adversarial packet) accuracy in localizing the positions of injected adversarial packets. 

\point{Evaluation Metrics}
For our evaluations of \tool, we use the following, most commonly adopted metrics in the state-of-the-art ML-based networked systems research \cite{mirsky2018kitsune,aqil2017jaal,marin2018rawpower,marin2019deepsec}. "Area Under of the Receiver Operating Characteristic Curve" (AUC-ROC) is our first metric of interest; it  captures the trade-off between the True Positive Rate (TPR) and the False Positive Rate (FPR) by considering a comprehensive set of reconstruction error thresholds, and is considered a comprehensive metric for evaluating binary classifiers. The higher the AUC-ROC, better the classifier. Complementary to AUC-ROC, EER is the point on ROC curve where the False Positive Rate is equal to the False Negative Rate; this metric is sometimes considered to be a more balanced metric to evaluate a classifier compared to AUC-ROC. The lower the ERR, the more accurate the detection. Lastly, \textit{Top-$N$ Hit Rate} is defined by the percentage of the connections where the context profiles with the $N$-highest reconstruction errors produced by \tool, intersect with  actual injected adversarial packets from among all tested connections. In other words, it measures the accuracy of the localization step in \tool's "localize-and-estimate" adversarial score generation approach, in terms of how well top candidates marked by \tool capture the real adversarial packets. With accurate localization, \tool can pinpoint the exact positions of adversarial packets for the purposes of forensic analyses. The higher the hit rate, the more accurate the localization.


\point{Detection Performance}
For evaluating \tool's detection accuracy, we must first inject the attacks into the benign traffic dataset to form its adversarial counterpart for the testing samples. As previously discussed, DPI evasion attacks all function by either injecting new packets, or modifying existing packets in a connection. However, the exact injection and modification methods differ from attack to attack. We therefore provide a breakdown/taxonomy of the evaluated attacks 
and the projects they were discovered from.

\begin{table*}[]
\small
\begin{tabular}{@{}|r|c|c|c|c|c|c|@{}}
\toprule
\begin{tabular}[c]{@{}r@{}}Results/\\ Approach\end{tabular} &
  \begin{tabular}[c]{@{}c@{}}Mean AUC-ROC\\ for \cite{wang2020symtcp}\end{tabular} &
  \begin{tabular}[c]{@{}c@{}}Mean EER\\ for \cite{wang2020symtcp}\end{tabular} &
  \begin{tabular}[c]{@{}c@{}}Mean AUC-ROC\\ for \cite{li2017lib}\end{tabular} &
  \begin{tabular}[c]{@{}c@{}}Mean EER\\ for \cite{li2017lib}\end{tabular} &
  \begin{tabular}[c]{@{}c@{}}Mean AUC-ROC\\ for \cite{bock2019geneva}\end{tabular} &
  \begin{tabular}[c]{@{}c@{}}Mean EER\\ for \cite{bock2019geneva}\end{tabular} \\ \midrule
\tool & \textbf{0.953}  & \textbf{0.072}   & \textbf{0.952}  & \textbf{0.082}   & \textbf{0.988}  & \textbf{0.024}   \\ \midrule
Baseline \#1         & 0.829 (-13.0\%) & 0.218 (+202.8\%) & 0.805 (-15.4\%) & 0.232 (+182.9\%) & 0.913 (-5.9\%)  & 0.133 (+454.2\%) \\ \midrule
Baseline \#2         & 0.501 (-47.4\%) & 0.501 (+595.8\%) & 0.500 (-47.5\%) & 0.500 (+509.8\%) & 0.491 (-50.3\%) & 0.504 (+2000\%)  \\ \bottomrule
\end{tabular}
\caption{Breakdown of average detection performance for strategies in \cite{wang2020symtcp,li2017lib,bock2019geneva}}
\label{tab:breakdown_per_work}
\end{table*}

\begin{figure*}
    \centering
    \includegraphics[width=\textwidth]{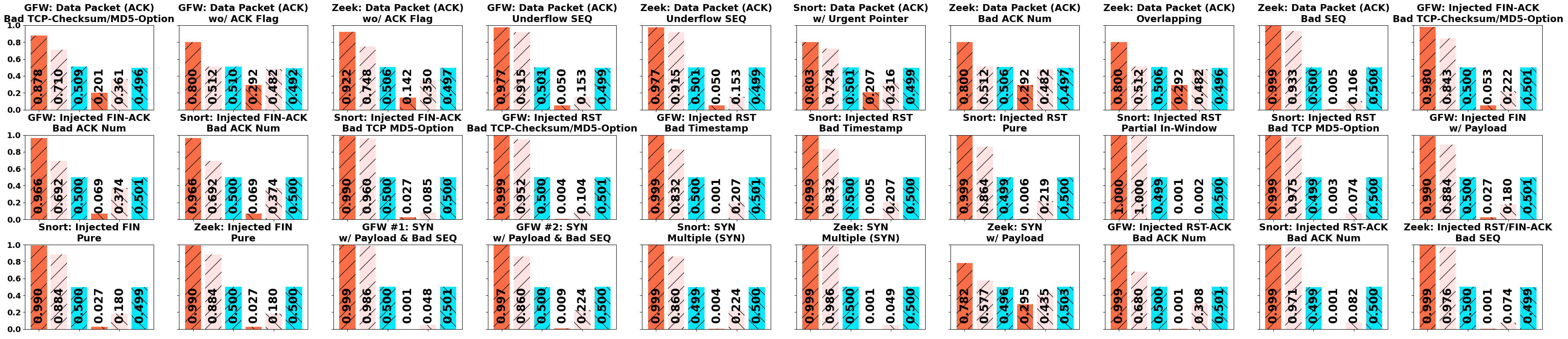}
    \caption{Per-strategy detection accuracy of \tool in detecting the different attacks (shown in title) from \cite{wang2020symtcp}}
    \label{fig:res_symtcp}
\end{figure*}

Figure \ref{fig:res_symtcp} shows the evaluation results of attacks presented in \cite{wang2020symtcp}. Since all of these attacks work at TCP layer by modifying the TCP header fields of injected or existing packets, they can be categorized based on (1) the type (i.e., TCP flags) of the packets that are injected or altered, 
and (2) the exact header modifications. In Figure \ref{fig:res_symtcp}, for each attack (i.e., the title of each bar plot), the first line is the key TCP flag (e.g., SYN) in the injected/altered packets, and the second line indicates how header fields are modified (e.g., Bad SEQ means changing the SEQ number to an invalid value).
As shown in Table \ref{tab:breakdown_per_work}, \tool outperforms both baselines in detecting evasion strategies from \cite{wang2020symtcp} by a significant margin.



\begin{figure*}
    \centering
    \includegraphics[width=\textwidth]{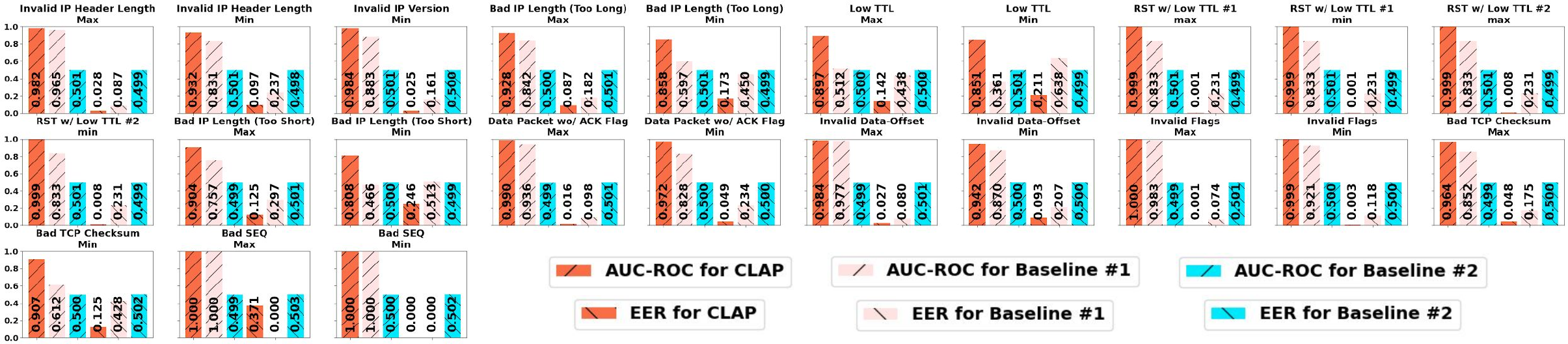}
    \caption{Per-strategy detection accuracy of \tool in detecting the different attacks (shown in title) from~\cite{li2017lib}}
    \label{fig:res_liberate}
\end{figure*}

Figure \ref{fig:res_liberate} shows the results relating to attacks proposed in \cite{li2017lib}. Unlike \cite{wang2020symtcp,bock2019geneva}, these attacks target DPI-based traffic classification systems 
(e.g., is it YouTube traffic or not?) by manipulating header fields across the TCP/IP layers in certain packets in a connection; these classifiers make decisions by examining an arbitrarily long subset of data packets called the matching packets, which are transferred after the initial TCP handshake. To evade the classifier, evasion packets are inserted in front of all of these matching packets. 
However, without knowing what classification possibilities the attacker is trying to evade we cannot know the exact number of evasion packets that are inserted (i.e., they vary depending on the content that requires evasion).  Hence,
we simulate two extreme cases wherein there is (a) a single matching packet \textit{min} and (b) where there are a \textit{max} = 5 matching packets as considered in the \cite{li2017lib} paper.
Note that these packets are sent after the connection transitions into the \texttt{ESTABLISHED} state.
With the above strategies we show the corresponding detection accuracies in Figure \ref{fig:res_liberate}.
Again, \tool outperforms both baselines in detecting evasion strategies from \cite{li2017lib} by a significant margin, as reported in Table \ref{tab:breakdown_per_work}



\begin{figure*}
    \centering
    \includegraphics[width=\textwidth]{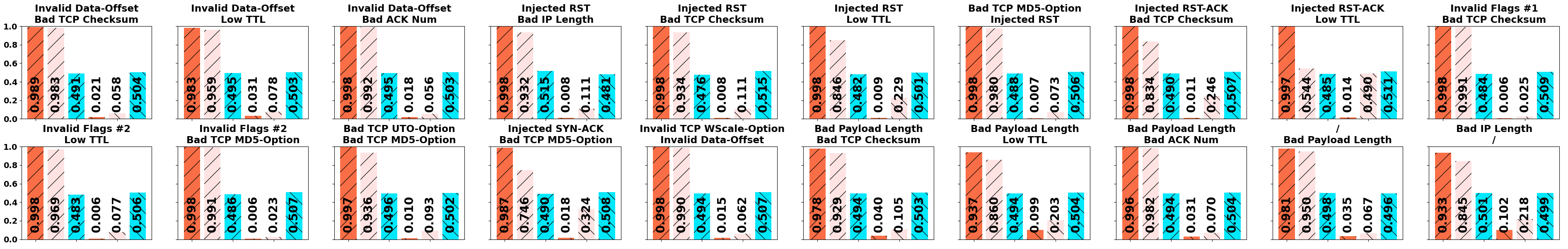}
    \caption{Per-strategy detection accuracy of \tool in detecting the different attacks (shown in title) from \cite{bock2019geneva}}
    \label{fig:res_geneva}
\end{figure*}

Lastly, Figure \ref{fig:res_symtcp} reports the detection accuracies with respect to attacks from \cite{bock2019geneva}. Unlike \cite{wang2020symtcp} and \cite{li2017lib}, these attacks (1) are executed/injected rather blindly and all data packets (i.e., packets transferred in \texttt{ESTABLISHED} TCP state after completing the handshake) are altered; and (2) consist up to 2 different modifications in one attack strategy. Therefore, in Figure \ref{fig:res_geneva}, when labeling each attack, the first and second lines describe the first and second modification type ("/" means only one modification for that strategy), respectively. We again observe significant gains in terms of detection performance, with \tool over the other baselines. 
Once again, \tool wins over both baselines in terms of the detection accuracy against attack strategies from \cite{bock2019geneva} by a considerable margin, as shown in Table \ref{tab:breakdown_per_work}.


\subpoint{Analysis}
From the detection performance results, alluded to above, across different evasion strategies, we have the following observations: (1)   \tool consistently outperforms both Baselines \#1 and \#2, in terms of all the considered metrics and across all attack strategies; (2)  all evaluation metrics indicate that \tool tends to perform exceptionally well on all evasion strategies that involve inter-packet context violations (e.g., Pure RST strategy from \cite{wang2020symtcp} with an AUC-ROC of 0.999, positioned at row \#2 column \#7 in Figure \ref{fig:res_symtcp}) and most strategies that involve intra-packet context violations (e.g., Invalid Data Offset/Bad TCP Checksum from \cite{bock2019geneva}, positioned at row \#1, column \#1 in Figure \ref{fig:res_geneva}); however, the performance is not as good on a small fraction of strategies involving intra-packet context violations (e.g., SYN w/ Payload attack from \cite{wang2020symtcp} with an AUC-ROC of 0.782, positioned at row \#3, column \#7 in Figure \ref{fig:res_symtcp} and, Low TTL attack (min) from \cite{li2017lib} with an AUC-ROC of 0.851, positioned at row \#1, column \#7 in Figure \ref{fig:res_liberate}). We believe that this is because, for these intra-packet context violations, even with amplification features, the quantum of the adversarial perturbation/modification imposed onto the packet is still considered too insignificant by the autoencoder in \tool, to be spotted. Note however that, even so, \tool  still (1) detects a large fraction of evasion attempts of these types (an AUC-ROC > 0.75); and (2) provides considerable improvements (> 30\%) over both baselines.

\begin{figure*}
    \centering
    \includegraphics[width=\textwidth]{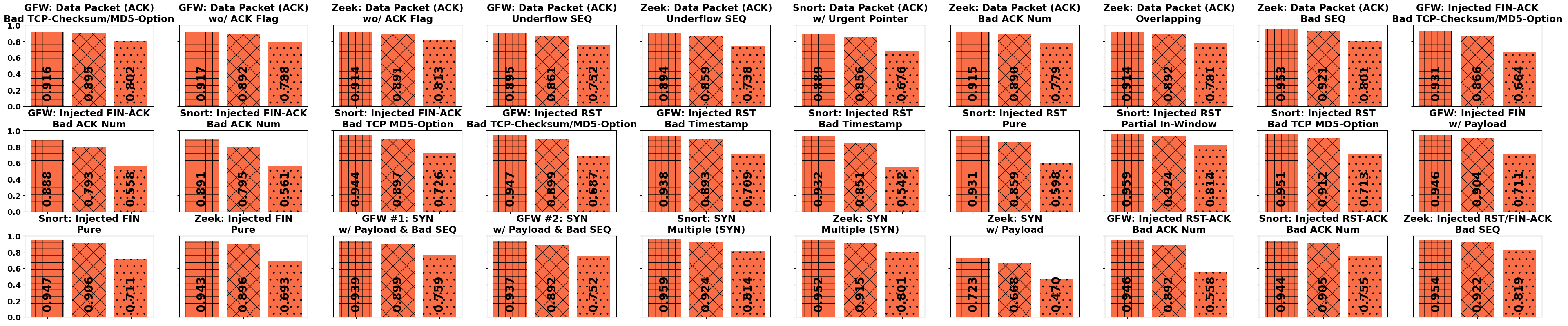}
    \caption{Per-strategy localization accuracy of \tool in detecting the different attacks (shown in title) from \cite{wang2020symtcp}}
    \label{fig:loc_res_symtcp}
\end{figure*}

\begin{figure*}
    \centering
    \includegraphics[width=\textwidth]{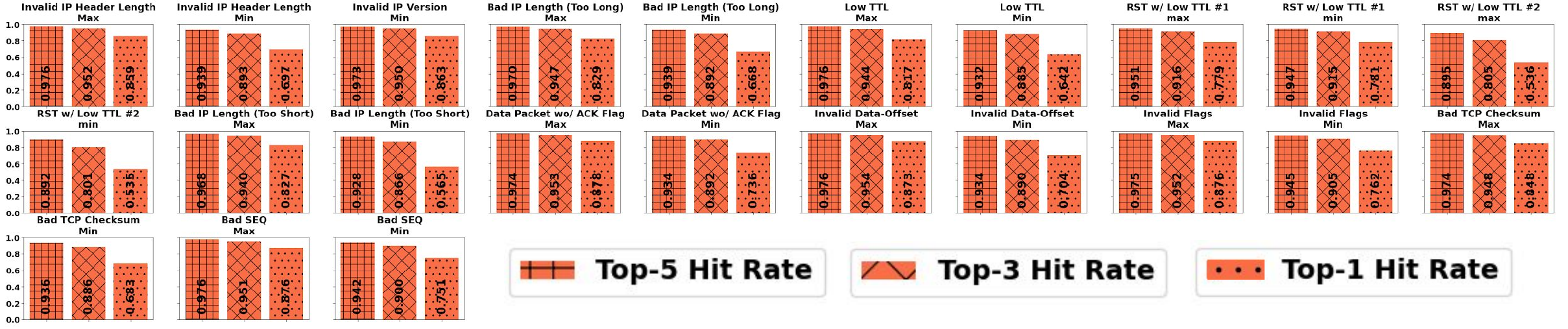}
    \caption{Per-strategy localization accuracy of \tool in detecting the different attacks from \cite{li2017lib}}
    \label{fig:loc_res_liberate}
\end{figure*}

\begin{figure*}
    \centering
    \includegraphics[width=\textwidth]{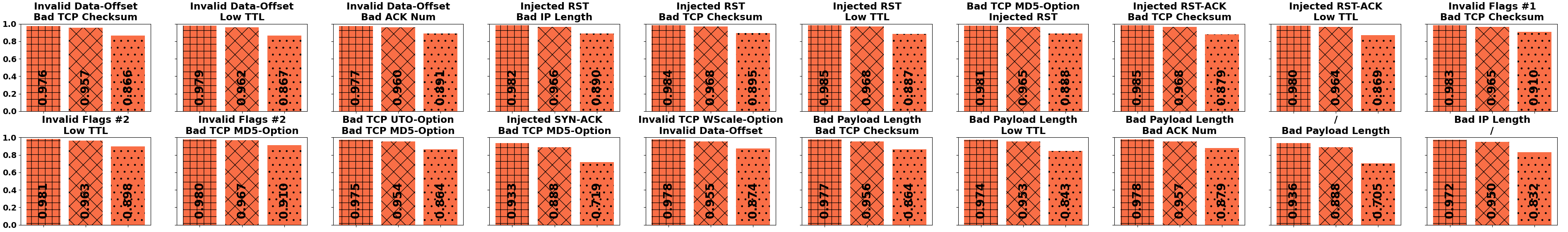}
    \caption{Per-strategy localization accuracy of \tool in detecting the different attacks from \cite{bock2019geneva}}
    \label{fig:loc_res_geneva}
\end{figure*}

\point{Localization Accuracy}
Next, we evaluate the accuracy of \tool in localizing the injected adversarial packets; this is the first step of the "localize-and-estimate" approach in computing the adversarial score for a given connection (see Section \ref{sec:adv_score}). 
Here, our goal is: (1) evaluating the localization ability of \tool, and (2) providing an analysis 
of the design choice made in \tool in association with the computation of adversarial score. We do not consider baseline approaches for localization evaluations, because neither of the baselines are sufficiently accurate in detecting adversarial packets in the first place and furthermore, do not consider the temporal dependencies across a train of packets (indicating their inadequacy in terms of localization abilities). As previously discussed, we use the Top-N hit rate as the metric to measure the localization accuracy. We report the Top-5, Top-3 and Top-1 rates in Figure \ref{fig:loc_res_symtcp}, \ref{fig:loc_res_liberate} and \ref{fig:loc_res_geneva} with the same categorizations adopted in the detection accuracy evaluations.  Specifically, the three bars in each plot from left to right correspond to Top-5, Top-3 and Top-1 rates, respectively.

\subpoint{Analysis}
Our general observations with regards to the localization results are as follows: (1) as one might expect, the accuracy varies when we require different levels of localization; specifically, achieving Top-1 (with an average accuracy across all strategies of 76.8\%) localization (as expected) is much more challenging than Top-5 (average accuracy 94.6\%) and Top-3 (average accuracy 91.0\%) localization. This in a way highlights the importance of our design choice with regards to using the ``mean across a sliding window'' surrounding the context profile with the maximum reconstruction error as the adversarial score described previously; (2) the better the localization, the better the detection performance, as more accurate localization yields richer and more directive context information for \tool to utilize.



\subsection{Case Studies}
\label{sec:eval_case}
Having provided a holistic picture of the performance of \tool across a large set of DPI evasion attacks, we 
now pick two concrete attacks to exemplify why \tool can detect them accurately.
Specifically, 
we pick an attack that is discovered due to violations of the inter-packet context {\color{black} and} an attack that is caught for violating the intra-packet context.
These case studies also serve to validate the core design of \tool. 

Note here that we have undertaken a careful analysis of all the attacks reported and are able to reason about them; we believe that the two attacks that we choose to magnify in this section are representative of the entire set. 
We also provide a summarized categorization of all \attknum evaluated strategies from \cite{wang2020symtcp,bock2019geneva,li2017lib} into the two context violation types, 
in Appendix \ref{tab:breakdown_per_context}. Recall that our Baseline \#1 approach does not account for inter-packet context since it only considers single-packet features  (gate weights from RNN are ignored). In other words, it only captures intra-packet context towards detecting DPI evasion attacks; thus,  any evasion strategy that exhibits disparity between \tool and Baseline \#1, is considered as one that invokes a inter-packet violation for evasion. Following this principle, if the AUC-ROC disparity between \tool and Baseline \#1 is greater than a chosen threshold $TH_{inter}$ (we pick 0.15 here), we categorize the attack as one invoking an inter-packet context violation; otherwise, the attack is considered as causing an intra-packet context violation. Based on this rule of thumb, we categorize 27 out of the 73 evasion attack strategies as primarily inter-packet violations, and 49 as primarily intra-packet violations. Note that this does not mean these two categories are strictly disjoint. While some strategies violate both contexts, we categorize them based on the main violation (manifested by the significant accuracy discrepancy).

\begin{table}[]
\small
\begin{tabular}{@{}|r|c|c|c|c|@{}}
\toprule
\multirow{2}{*}{\begin{tabular}[c]{@{}r@{}}Category/\\ Metric\end{tabular}} & \multicolumn{2}{c|}{\begin{tabular}[c]{@{}c@{}}Inter-packet\\ Context Violation\\ (24 Strategies)\end{tabular}} & \multicolumn{2}{c|}{\begin{tabular}[c]{@{}c@{}}Intra-packet\\ Context Violation\\ (49 Strategies)\end{tabular}} \\ \cmidrule(l){2-5} 
 & \tool & Basline \#1 & \tool & Baseline \#1 \\ \midrule
\begin{tabular}[c]{@{}r@{}}Mean\\ AUC-ROC\end{tabular} & \textbf{\begin{tabular}[c]{@{}c@{}}0.925\\ (+37.6\%)\end{tabular}} & 0.672 & \textbf{\begin{tabular}[c]{@{}c@{}}0.980\\ (+6.2\%)\end{tabular}} & 0.923 \\ \midrule
\begin{tabular}[c]{@{}r@{}}Mean\\ EER\end{tabular} & \textbf{\begin{tabular}[c]{@{}c@{}}0.109\\ (-70.0\%)\end{tabular}} & 0.364 & \textbf{\begin{tabular}[c]{@{}c@{}}0.039\\ (-68.3\%)\end{tabular}} & 0.123 \\ \bottomrule
\end{tabular}
\caption{Breakdown of inter vs intra-packet violation detections}
\label{tab:context_breakdown}
\end{table}

\point{Inter-packet context violation}
Recall that the inter-packet context refers to the inter-relationships across different packets in a connection. Among all evaluated attacks, there are many that evade DPI detection by injecting manipulated packets that cause a behavioural discrepancy between the DPI and endhost only when the TCP state machine is in a specific state. For example, the RST with Bad Timestamp strategy
from \cite{wang2020symtcp} injects a RST packet with an invalid TCP timestamp option value, specifically when the target connection is in the \texttt{SYN\_RECV} state. Since this evasion packet is then only  dropped by endhost, but accepted by target DPIs (e.g. Zeek and GFW), it tricks the DPI into disengaging the monitoring subsequently. The end host, which is in exactly \texttt{SYN\_RECV} TCP state, receives the follow-up packets that effectively bypass DPI detection. We determine that this strategy primarily violates the inter-packet context because it (1) it is successful only when the current TCP state is in a specific state (i.e., a Bad Timestamp RST packet would not cause behavioural discrepancy between DPI and endhost in \texttt{ESTABLISHED} state) and, (2) the validity of the TCP timestamp option is determined by the timestamps in previous packets. In other words, in order to detect this strategy accurately, \tool must utilize the inter-packet context/relationship. It asserts if a given packet (1) bears a bad timestamp relative to  previous packets, and (2) if the current TCP state is \texttt{SYN\_RECV}. As expected, in this case, Figure \ref{tab:context_breakdown} shows that \tool performs much better than Baseline \#1 (> 35\%).

\point{Intra-packet context violation}
Recall that the intra-packet context captures the relationships across different header fields in the same packet. Several evaluated attacks inject shadow packets in front of legal data packets. In these shadow packets, certain TCP/IP header field values are altered such that they are rejected by end point's rigorous TCP implementation but accepted by DPI's simplified version. This way, the data packets that follow the injected shadow packets are cloaked and evade DPI detection. For example, the \textit{Invalid IP Version} attack from \cite{li2017lib} injects packets with an incorrect IP Version header value (e.g 5 as there is no IPv5) to trigger the discrepancy between the DPI and endhost. \tool detects this attack by learning the intra-packet context i.e., legitimate packets should not have a IP Version of 5; it is thus able to flag any packet that violates this requirement. The aggregate accuracies in Table \ref{tab:context_breakdown} show that \tool performs considerably better than Baseline \#1.

\subsection{Runtime Overhead Analysis}
\label{sec:eval_perf}

Finally, we evaluate how efficiently \tool processes network traffic streams with our current implementation. Given that the training phase of \tool is  completely offline,
the critical overhead 
would be its runtime overhead, i.e., the model processing efficiency of \tool's testing phase (Stage (d)). For reference, we also measure the runtime model inference overhead of the open-source version released by \cite{mirsky2018kitsune}, which is considered to be the state-of-the-art autoencoder-based IDS, and show that \tool achieves a significantly higher inference/processing speed. We use the default hyper-parameters to train the model from \cite{mirsky2018kitsune}, as detailed in Table \ref{tab:hyperparameters}.

\point{Setup}
We run the pipelines of both \tool and \cite{mirsky2018kitsune} on a desktop with Intel Xeon E3-1225 Processor (3.2Ghz, 4 cores), 20GB RAM and disabled GPU support. 
We constrain both pipelines to only use one logical core for fair comparison. We note that \cite{mirsky2018kitsune} claims a higher processing throughput using a C++ implementation, but find that its released Python version \cite{kitsunerepo}
is much slower. Furthermore, it is fair to compare our Python-based prototype implementation with its Python-based baseline counterpart -- we leave the task of re-implementing CLAP in more efficient languages (e.g. C/C++) for improving its overhead performance as future work. Our test adversarial traffic corpus consists of 92,262 packets and 6,424 connections.

\point{Throughput}
We compare the model processing throughputs of \tool and \cite{mirsky2018kitsune} in Table \ref{tab:throughput}. We see that \tool outperforms \cite{mirsky2018kitsune} by a margin of $\approx$ 50\%,
while detecting DPI evasion attacks with much higher accuracies, as shown in previous subsections. 
However, we acknowledge that \tool is not designed for detecting other network attacks, that are captured by \cite{mirsky2018kitsune}.
We believe that the performance disparity is because \cite{mirsky2018kitsune} uses an ensemble (forest) of 10 small autoencoders to process subsets of different features pertaining to different attacks (details in Table \ref{tab:hyperparameters}); it merges the separate reconstruction errors into an aggregate error with another autoencoder, making it overall a heavy procedure i.e., it cannot process packets as fast as \tool (with only one autoencoder).


\begin{table}
\small
\begin{tabular}{@{}|r|c|c|@{}}
\toprule
\begin{tabular}[c]{@{}r@{}}Model/\\ Metric\end{tabular} & \tool & \cite{mirsky2018kitsune} \\ \midrule
Packets/Second & \textbf{2,162.2 (+49.7\%)} & 1,444.5 \\ \midrule
Connections/Second & \textbf{97.0 (+49.7\%)} & 64.8 \\ \bottomrule
\end{tabular}
\caption{Model processing throughput}
\label{tab:throughput}
\end{table}




\section{Discussion}
\point{Ethical Implications}
In recent years, DPI evasion attacks have been considered as a means to bypass censorship systems \cite{wang2020symtcp,bock2019geneva,li2017lib,wang2017your}. This is because, censors are essentially DPI middleboxes.
We acknowledge that malicious DPI evasion attacks cannot be fully separated from censorship circumvention practices (since they rely on DPI evasion attacks). 
However, while \tool can inherently enable censorship i.e., potentially expose censorship circumvention,  we do not advocate censorship and our work is completely inspired by research questions.

\point{Feature Completeness}
Currently, \tool covers all non-optional, non-tuple-related header fields. One might wonder if 
new evasion strategies that are outside our current feature set are viable. Since \tool currently does not consider (1) uncommon IP and TCP option fields that are variable-sized (i.e., hard to encode as fix-sized features); and (2) payloads as they are unstructured data (i.e., with no fixed formats), we envision that these are the most likely surfaces that future emerging evasions can manipulate. To tackle this challenge, we plan to explore the remaining headers (TCP/IP options) and payload contents, possibly via embedding techniques \cite{wu2020comprehensive} that can map unstructured/variable-sized inputs to fix-sized vectors in future work.

\point{Generalizability}
In this work, we focus on evasions via attacks of TCP/IP protocols as these protocols are the most popular/common parts of most network protocol suites used today, and thus attract the vast majority of evasion efforts. However, we believe that the pipeline and core design of \tool can be easily transferred and applied to other stateful protocols (e.g., HTTP) with  minimum effort. As long as one can define the internal states inside the protocol implementation, \tool is expected to learn packet contexts that help enable adversarial evasions. 

\point{Deployability}
As shown in Section \ref{sec:eval_perf}, the model inference time for \tool is considerably faster than the prototype released by a state-of-the-art ML-based IDS, making \tool 
potentially deployable on middleboxes on low-workload networks. We acknowledge that compared to signature-based general-purpose IDSs such as Zeek, our overall processing speed is considerably slower, but argue that (1) we target specifically DPI evasion attacks that cannot be captured by general-purpose IDS; (2) a re-implementation of \tool in more efficient languages (e.g. C/C++) can greatly help reduce its overhead (this is left to future work); and (3) the rising trend of GPU-based ML acceleration can also be of great help in boosting the runtime performance of \tool, which is again, part of our future plan.

We acknowledge that even though we have demonstrated that \tool comes with very low error rates in both detection and localization tasks in Section \ref{sec:eval}, there can still be a number of false alarms that might be generated. However, we emphasize that \tool can be tuned using  a pre-defined adversarial score threshold (see Section \ref{sec:adv_score}), which would allow  the deployers of \tool the flexibility to freely choose the desired trade-off between true positive and false positive rates. In addition, one can down sample the alarms generated by \tool, and only selectively inspect/analyze those packets that have the highest associated adversarial scores. We believe these two strategies can significantly control the number/ratio of false alarms.

\point{Attacker Adaptation}
We are aware of the emerging research that exploits inherent vulnerabilities against ML models such that specifically crafted inputs, known as adversarial examples, can bypass ML models with minimum differences compared to its benign counterparts. We envision the possibility that attackers could generate such examples (packets) to hide from \tool's detection. However, we point out that most of the adversarial example attacks
from the adversarial ML community, consider end-to-end models
so that it is rather easy to
obtain the gradients from such end-to-end models to guide the
generation of adversarial examples. In CLAP our design
choices of using a multi-step pipeline (i.e., RNN + autoencoder)
makes it extremely challenging, if possible at all.
\section{Conclusions}
In this paper, we design and implement a framework (called \tool) for detecting DPI evasion attacks that have emerged recently. The key observation that drives the design of \tool is that these attacks often violate either legitimate relationships between the headers within a packet, or relationships across headers in a train of packets. We construct a packet context-profile that captures these relationships and train a set of appropriate ML models to learn the legitimate (benign) distributions of such profiles. During test time, \tool checks if encountered packets conform with these distributions and flags those that do not as evasion attempts. Our comprehensive evaluations on a large variety of attacks from three different recent efforts show that \tool (a) is extremely effective in detecting evasion patterns, and (b) significantly outperforms ML methods agnostic to packet context-profiles.

\section*{Acknowledgement}

This research was sponsored by the
U.S. Army Combat Capabilities Development Command Army
Research Laboratory and was accomplished under Cooperative
Agreement Number W911NF-13-2-0045 (ARL Cyber Security
CRA). The views and conclusions contained in this document
are those of the authors and should not be interpreted as representing the official policies, either expressed or implied, of the Combat Capabilities Development Command Army Research
Laboratory or the U.S. Government. The U.S. Government is
authorized to reproduce and distribute reprints for Government
purposes notwithstanding any copyright notation here on.
We also thank the anonymous reviewers and our anonymous shepherd for their comments and suggestions that helped significantly improve our paper.

\bibliographystyle{ACM-Reference-Format}
\bibliography{reference}

\appendix
\onecolumn
\section*{Appendix}
\label{sec:appendix}

\section{MAWI Traffic Dataset Statistics}
As described in Section \ref{sec:eval_setup}, here we list the basic statistics of the dataset used in our evaluations in Table \ref{tab:dataset}.

\section{RNN Prediction Accuracy}
We present the per-label accuracy breakdown along with the number of samples (packets) in Table \ref{tab:rnn_accuracy}. Overall, the RNN model in \tool achieves an accuracy of 0.995 on its testing set.

\section{Feature Set}
Here we list the features we used in building context profiles in Table \ref{tab:feature_set}. Note that, as previously described, only TCP and IP layer header features (i.e., \#1-\#32) are used for training the RNN model.

\section{Per-context Categorization of Evasion Strategies}
We categorize the \attknum evasion strategies that were evaluated in Section \ref{sec:eval_eff} in Table \ref{tab:breakdown_per_context} according to the context they primarily violate, using the categorization scheme proposed previously.

\section{Model Hyper-parameters}
We summarize the hyper-parameters used in training the RNN, autoencooder models in \tool, and the autoencoder models used in Baseline \#1 and \#2 for our evaluations, in Table \ref{tab:hyperparameters}.

\begin{table}[htb]
\tiny
\begin{tabular}{@{}|c|c|c|c|@{}}
\toprule
\multicolumn{4}{|c|}{\textbf{Original Trafffc Archive}} \\ \midrule
\begin{tabular}[c]{@{}c@{}}Capture\\ Timestamp\end{tabular} & 04/07/2020 14:00 JPT & \# Packets & 111,851,572 \\ \midrule
\begin{tabular}[c]{@{}c@{}}\# TCP/IPv4\\ Packets\end{tabular} & 51,692,562 & \multicolumn{2}{c|}{/} \\ \midrule
\multicolumn{4}{|c|}{\textbf{Sampled (Used) Dataset}} \\ \midrule
\begin{tabular}[c]{@{}c@{}}\# TCP/IPv4\\ Packets\end{tabular} & 540,353 & \begin{tabular}[c]{@{}c@{}}\# TCP/IPv4\\ Connections\end{tabular} & 37,622 \\ \midrule
\begin{tabular}[c]{@{}c@{}}\# TCP/IPv4\\ Packets (Training)\end{tabular} & 448,091 & \begin{tabular}[c]{@{}c@{}}TCP/IPv4\\ Connections (Training)\end{tabular} & 31,198 \\ \midrule
\begin{tabular}[c]{@{}c@{}}\# TCP/IPv4\\ Packets (Testing)\end{tabular} & 92,262 & \begin{tabular}[c]{@{}c@{}}\# TCP/IPv4\\ Connections (Testing)\end{tabular} & 6,424 \\ \bottomrule
\end{tabular}
\caption{Statistics of used MAWI dataset}
\label{tab:dataset}
\end{table}

\begin{table*}[htb]
\tiny
\begin{tabular}{@{}|r|c|c|c|c|c|c|c|c|@{}}
\toprule
\begin{tabular}[c]{@{}r@{}}TCP State/\\ Packet Window Classification\end{tabular} & SYN\_SENT & SYN\_RECV & ESTABLISHED & FIN\_WAIT & CLOSE\_WAIT & LAST\_ACK & TIME\_WAIT & CLOSE \\ \midrule
In-Window & \begin{tabular}[c]{@{}c@{}}0.999678\\ (6166)\end{tabular} & \begin{tabular}[c]{@{}c@{}}1.0\\ (18906)\end{tabular} & \begin{tabular}[c]{@{}c@{}}0.994733\\ (47664)\end{tabular} & \begin{tabular}[c]{@{}c@{}}0.996628\\ (2966)\end{tabular} & \begin{tabular}[c]{@{}c@{}}0.988205\\ (3117)\end{tabular} & \begin{tabular}[c]{@{}c@{}}0.993641\\ (2988)\end{tabular} & \begin{tabular}[c]{@{}c@{}}0.992513\\ (2805)\end{tabular} & \begin{tabular}[c]{@{}c@{}}0.987467\\ (2314)\end{tabular} \\ \midrule
Out-of-Window & \begin{tabular}[c]{@{}c@{}}0.0\\ (2)\end{tabular} & / & \begin{tabular}[c]{@{}c@{}}0.953246\\ (1155)\end{tabular} & \begin{tabular}[c]{@{}c@{}}0.911764\\ (34)\end{tabular} & \begin{tabular}[c]{@{}c@{}}0.694444\\ (36)\end{tabular} & \begin{tabular}[c]{@{}c@{}}0.814818\\ (27)\end{tabular} & \begin{tabular}[c]{@{}c@{}}0.95\\ (20)\end{tabular} & \begin{tabular}[c]{@{}c@{}}0.956521\\ (23)\end{tabular} \\ \bottomrule
\end{tabular}
\caption{Per-label breakdown of RNN accuracy}
\label{tab:rnn_accuracy}
\end{table*}

\begin{table*}[htb]
\tiny
\begin{tabular}{@{}|r|c|c|c|c|c|c|c|c|c|c|@{}}
\toprule
\multicolumn{1}{|c|}{} & \multicolumn{4}{c|}{RNN (GRU-based) in CLAP} & \multicolumn{6}{c|}{Autoencoder in CLAP} \\ \midrule
Model Parameter & \# Layer(s) & Input Size & \begin{tabular}[c]{@{}c@{}}Hidden Size\\ (Gate Size)\end{tabular} & \# Epochs & \# Layer(s) & Input Size & \multicolumn{2}{c|}{\begin{tabular}[c]{@{}c@{}}Length of Context \\ Profile Stacking\end{tabular}} & \begin{tabular}[c]{@{}c@{}}Bottleneck Layer\\ Size\end{tabular} & \# Epochs \\ \midrule
Value & 1 & 32 & 32 & 30 & 7 & 345 & \multicolumn{2}{c|}{3} & 40 & 1,000 \\ \midrule
 & \multicolumn{4}{c|}{Autoencoder in Baseline \#1} & \multicolumn{6}{c|}{Ensembled Autoencoders in Baseline \#2} \\ \midrule
Model Parameter & \# Layer(s) & Input Size & \begin{tabular}[c]{@{}c@{}}Bottleneck Layer\\ Size\end{tabular} & \# Epochs & \# Layer(s) & \begin{tabular}[c]{@{}c@{}}Ensemble Size\\ (\# Autoencoders)\end{tabular} & \begin{tabular}[c]{@{}c@{}}Total Input\\ Size\end{tabular} & \begin{tabular}[c]{@{}c@{}}Average Input Size\\ (Per Autoencoder)\end{tabular} & \begin{tabular}[c]{@{}c@{}}Average Bottleneck \\ Layer Size\end{tabular} & \# Epochs \\ \midrule
Value & 3 & 51 & 5 & 1,000 & 1 & 16 & 100 & 6.25 & 4.68 & 1 \\ \bottomrule
\end{tabular}
\caption{Hyper-parameters used in the paper}
\label{tab:hyperparameters}
\end{table*}

\begin{table*}[htb]
\tiny
\begin{tabular}{@{}|c|c|c|c|c|c|c|c|c|c|c|c|@{}}
\toprule
Index & Type & Sementic & Index & Type & Sementic & Index & Type & Sementic & Index & Type & Sementic \\ \midrule
\multicolumn{3}{|c|}{\textbf{TCP Layer Features}} & 17 & Integer & Payload Length & \multicolumn{3}{c|}{\textbf{IP Layer Features}} & \multicolumn{3}{c|}{\textbf{\begin{tabular}[c]{@{}c@{}}Amplification Features\\ (not included for training RNN)\end{tabular}}} \\ \midrule
1 & Binary & Packet direction & 18 & Integer & \begin{tabular}[c]{@{}c@{}}Option: \\ Maximum Segment Size\end{tabular} & 26 & Integer & Length & 33-45 & Binary & \begin{tabular}[c]{@{}c@{}}Out-of-Range indicators\\ for numeric TCP features\end{tabular} \\ \midrule
2 & Integer & \begin{tabular}[c]{@{}c@{}}SEQ number \\ (incremental)\end{tabular} & 19 & Integer & \begin{tabular}[c]{@{}c@{}}Option: Timestamp Value\\ (TSVal)\end{tabular} & 27 & Integer & Time-To-Live & 46-50 & Binary & \begin{tabular}[c]{@{}c@{}}Out-of-Range indicators\\ for numeric IP features\end{tabular} \\ \midrule
3 & Integer & \begin{tabular}[c]{@{}c@{}}ACK number \\ (incremental)\end{tabular} & 20 & Integer & \begin{tabular}[c]{@{}c@{}}Option: Timestamp\\ Echo Reply (TSecr)\end{tabular} & 28 & Integer & Header Length & 51 & Binary & \begin{tabular}[c]{@{}c@{}}TCP Payload Length correctness\\ (\#17 = \#26 - \#28 - \#4)\end{tabular} \\ \midrule
4 & Integer & Data Offset & 21 & Integer & Option: Window Scale & 29 & Binary & Checksum validity & \multicolumn{3}{c|}{\textbf{Gate Weights from GRU}} \\ \midrule
5-13 & Categorical & \begin{tabular}[c]{@{}c@{}}Flags \\ (one-hot encoded)\end{tabular} & 22 & Integer & Option: User Timeout & 30 & Integer & IP Version & 52-83 & Float & Update Gates \\ \midrule
14 & Integer & Window Size & 23 & Binary & Option: MD5 Header Validity & 31 & Integer & Type of Service & 84-115 & Float & Reset Gates \\ \midrule
15 & Binary & Checksum vadility & 24 & Integer & TCP Timestamp & 32 & Binary & \begin{tabular}[c]{@{}c@{}}Existence of non-standard\\ IP options\end{tabular} & \multicolumn{3}{c|}{\multirow{2}{*}{}} \\ \cmidrule(r){1-9}
16 & Integer & Urgent Pointer & 25 & Integer & Frame Timestamp & \multicolumn{3}{c|}{} & \multicolumn{3}{c|}{} \\ \bottomrule
\end{tabular}
\caption{List of features in context profile}
\label{tab:feature_set}
\end{table*}

\begin{table*}[htb]
\tiny
\begin{tabular}{@{}|c|c|c|c|c|c|@{}}
\toprule
From & Strategy Name & From & Strategy Name & From & Strategy Name \\ \midrule
\multicolumn{2}{|c|}{\textbf{Inter-packet Context Violation}} & \multirow{16}{*}{\cite{wang2020symtcp}} & Zeek: Data Packet (ACK) Bad SEQ & \multirow{10}{*}{\cite{li2017lib}} & Bad SEQ (Min) \\ \cmidrule(r){1-2} \cmidrule(lr){4-4} \cmidrule(l){6-6} 
\multirow{11}{*}{\cite{wang2020symtcp}} & GFW: Data Packet (ACK) Bad TCP-Checksum/MD5-Option &  & GFW: Injected FIN-ACK Bad TCP-Checksum/MD5-Option &  & Data Packet wo/ ACK Flag (Max) \\ \cmidrule(lr){2-2} \cmidrule(lr){4-4} \cmidrule(l){6-6} 
 & GFW: Data Packet (ACK) wo/ ACK Flag &  & Snort: Injected FIN-ACK Bad TCP MD5-Option &  & Data Packet wo/ ACK Flag (Min) \\ \cmidrule(lr){2-2} \cmidrule(lr){4-4} \cmidrule(l){6-6} 
 & Zeek: Data Packet (ACK) wo/ ACK Flag &  & GFW: Injected RST Bad TCP-Checksum/MD5-Option &  & Invalid Data-Offset (Max) \\ \cmidrule(lr){2-2} \cmidrule(lr){4-4} \cmidrule(l){6-6} 
 & Zeek: Data Packet (ACK) Bad ACK Num &  & Snort: Injected RST Pure &  & Invalid Data-Offset (Min) \\ \cmidrule(lr){2-2} \cmidrule(lr){4-4} \cmidrule(l){6-6} 
 & Zeek: Data Packet (ACK) Overlapping &  & Snort: Injected RST Partial In-Window &  & Invalid Flags (Max) \\ \cmidrule(lr){2-2} \cmidrule(lr){4-4} \cmidrule(l){6-6} 
 & GFW: Injected FIN-ACK Bad ACK Num &  & Snort: Injected RST Bad TCP MD5-Option &  & Invalid Flags (Min) \\ \cmidrule(lr){2-2} \cmidrule(lr){4-4} \cmidrule(l){6-6} 
 & Snort: Injected FIN-ACK Bad ACK Num &  & GFW: Injected FIN w/ Payload &  & Bad TCP Checksum (Max) \\ \cmidrule(lr){2-2} \cmidrule(lr){4-4} \cmidrule(l){6-6} 
 & GFW: Injected RST Bad Timestamp &  & Snort: Injected FIN Pure &  & Bad SEQ (Max) \\ \cmidrule(lr){2-2} \cmidrule(lr){4-4} \cmidrule(l){6-6} 
 & Snort: Injected RST Bad Timestamp &  & Zeek: Injected FIN Pure &  & Bad SEQ (Min) \\ \cmidrule(lr){2-2} \cmidrule(l){4-6} 
 & Zeek: SYN w/ Payload &  & GFW \#1: SYN w/ Payload \& Bad SEQ & \multirow{16}{*}{\cite{bock2019geneva}} & \begin{tabular}[c]{@{}c@{}}Invalid Data-Offset \\ Bad TCP Checksum\end{tabular} \\ \cmidrule(lr){2-2} \cmidrule(lr){4-4} \cmidrule(l){6-6} 
 & GFW: Injected RST-ACK Bad ACK Num &  & GFW \#2: SYN w/ Payload \& Bad SEQ &  & \begin{tabular}[c]{@{}c@{}}Invalid Data-Offset \\ Low TTL\end{tabular} \\ \cmidrule(r){1-2} \cmidrule(lr){4-4} \cmidrule(l){6-6} 
\multirow{9}{*}{\cite{li2017lib}} & Bad IP Length (Too Long) (Min) &  & Snort: SYN Multiple (SYN) &  & \begin{tabular}[c]{@{}c@{}}Invalid Data-Offset \\ Bad ACK Num\end{tabular} \\ \cmidrule(lr){2-2} \cmidrule(lr){4-4} \cmidrule(l){6-6} 
 & Low TTL (Max) &  & Zeek: SYN Multiple (SYN) &  & \begin{tabular}[c]{@{}c@{}}Injected RST \\ Bad IP Length\end{tabular} \\ \cmidrule(lr){2-2} \cmidrule(lr){4-4} \cmidrule(l){6-6} 
 & Low TTL (Min) &  & Snort: Injected RST-ACK Bad ACK Num &  & \begin{tabular}[c]{@{}c@{}}Injected RST Bad \\ TCP Checksum\end{tabular} \\ \cmidrule(lr){2-2} \cmidrule(lr){4-4} \cmidrule(l){6-6} 
 & RST w/ Low TTL \#1 (Max) &  & Zeek: Injected RST/FIN-ACK Bad SEQ &  & \begin{tabular}[c]{@{}c@{}}Bad TCP MD5-Option \\ Injected RST\end{tabular} \\ \cmidrule(lr){2-4} \cmidrule(l){6-6} 
 & RST w/ Low TTL \#1 (Min) & \multirow{13}{*}{\cite{li2017lib}} & Invalid IP Header Length (Max) &  & \begin{tabular}[c]{@{}c@{}}Invalid Flags \#1 \\ Bad TCP Checksum\end{tabular} \\ \cmidrule(lr){2-2} \cmidrule(lr){4-4} \cmidrule(l){6-6} 
 & RST w/ Low TTL \#2  (Max) &  & Invalid IP Header Length (Min) &  & \begin{tabular}[c]{@{}c@{}}Invalid Flags \#2 \\ Low TTL\end{tabular} \\ \cmidrule(lr){2-2} \cmidrule(lr){4-4} \cmidrule(l){6-6} 
 & RST w/ Low TTL \#2 (Min) &  & Invalid IP Version (Min) &  & \begin{tabular}[c]{@{}c@{}}Invalid Flags \#2 \\ Bad TCP MD5-Option\end{tabular} \\ \cmidrule(lr){2-2} \cmidrule(lr){4-4} \cmidrule(l){6-6} 
 & Bad IP Length (Too Short) (Min) &  & Bad IP Length (Too Long) (Max) &  & \begin{tabular}[c]{@{}c@{}}Bad TCP UTO-Option \\ Bad TCP MD5-Option\end{tabular} \\ \cmidrule(lr){2-2} \cmidrule(lr){4-4} \cmidrule(l){6-6} 
 & Bad TCP Checksum (Min) &  & Bad IP Length (Too Short) (Max) &  & \begin{tabular}[c]{@{}c@{}}Invalid TCP WScale-Option \\ Invalid Data-Offset\end{tabular} \\ \cmidrule(r){1-2} \cmidrule(lr){4-4} \cmidrule(l){6-6} 
\multirow{4}{*}{\cite{bock2019geneva}} & \begin{tabular}[c]{@{}c@{}}Injected RST \\ Low TTL\end{tabular} &  & Data Packet wo/ ACK Flag (Max) &  & \begin{tabular}[c]{@{}c@{}}Bad Payload Length \\ Bad TCP Checksum\end{tabular} \\ \cmidrule(lr){2-2} \cmidrule(lr){4-4} \cmidrule(l){6-6} 
 & \begin{tabular}[c]{@{}c@{}}Injected RST-ACK \\ Bad TCP Checksum\end{tabular} &  & Data Packet wo/ ACK Flag (Min) &  & \begin{tabular}[c]{@{}c@{}}Bad Payload Length \\ Low TTL\end{tabular} \\ \cmidrule(lr){2-2} \cmidrule(lr){4-4} \cmidrule(l){6-6} 
 & \begin{tabular}[c]{@{}c@{}}Injected RST-ACK \\ Low TTL\end{tabular} &  & Invalid Data-Offset (Max) &  & \begin{tabular}[c]{@{}c@{}}Bad Payload Length \\ Bad ACK Num\end{tabular} \\ \cmidrule(lr){2-2} \cmidrule(lr){4-4} \cmidrule(l){6-6} 
 & \begin{tabular}[c]{@{}c@{}}Injected SYN-ACK \\ Bad TCP MD5-Option\end{tabular} &  & Invalid Data-Offset (Min) &  & \begin{tabular}[c]{@{}c@{}}/ \\ Bad Payload Length\end{tabular} \\ \cmidrule(r){1-2} \cmidrule(lr){4-4} \cmidrule(l){6-6} 
\multicolumn{2}{|c|}{\textbf{Intra-packet Context Violation}} &  & Invalid Flags (Max) &  & \begin{tabular}[c]{@{}c@{}}Bad IP Length \\ /\end{tabular} \\ \cmidrule(r){1-2} \cmidrule(l){4-6} 
\multirow{3}{*}{\cite{wang2020symtcp}} & GFW: Data Packet (ACK) Underflow SEQ &  & Invalid Flags (Min) & \multicolumn{2}{c|}{\multirow{3}{*}{}} \\ \cmidrule(lr){2-2} \cmidrule(lr){4-4}
 & Zeek: Data Packet (ACK) Underflow SEQ &  & Bad TCP Checksum (Max) & \multicolumn{2}{c|}{} \\ \cmidrule(lr){2-2} \cmidrule(lr){4-4}
 & Snort: Data Packet (ACK) w/ Urgent Pointer &  & Bad SEQ (Max) & \multicolumn{2}{c|}{} \\ \bottomrule
\end{tabular}
\caption{Per-context categorization of evasion strategies from \cite{wang2020symtcp,li2017lib,bock2019geneva} (with $TH_{inter}=0.15$)}
\label{tab:breakdown_per_context}
\end{table*}

\end{document}